\documentclass[11pt,a4paper]{article}
\pdfoutput=1
\usepackage[hmargin=3.0cm,vmargin=3.0cm]{geometry}

\newcommand{\jacob}{\mathbf{J}}

\newcommand{\bmx}[0]{\begin{bmatrix}}
\newcommand{\emx}[0]{\end{bmatrix}}

\newcommand{\vect}[1]{\mathbf{#1}}
\newcommand{\vects}[1]{\boldsymbol{#1}}
\newcommand{\matr}[1]{\mathbf{#1}}

\newcommand{\va}[0]{\vect{a}}
\newcommand{\vb}[0]{\vect{b}}

\newcommand{\vv}[0]{\vect{v}}

\newcommand{\vw}[0]{\vect{w}}

\newcommand{\vf}[0]{\vect{f}}
\newcommand{\vy}[0]{\vect{y}}

\newcommand{\vd}[0]{\vect{d}}

\newcommand{\vu}[0]{\vect{u}}

\newcommand{\vr}[0]{\vect{r}}
\newcommand{\vF}[0]{\vect{F}}
\newcommand{\mW}[0]{\matr{W}}

\newcommand{\mX}[0]{\matr{X}}

\newcommand{\mU}[0]{\matr{U}}
\newcommand{\mV}[0]{\matr{V}}
\newcommand{\mA}{\matr{A}}
\newcommand{\mB}{\matr{B}}
\newcommand{\mD}{\matr{D}}

\newcommand{\mI}{\matr{I}}
\newcommand{\mJ}{\matr{J}}

\newcommand{\mK}{\matr{K}}
\newcommand{\mP}{\matr{P}}

\newcommand{\TT}[0]{{\vects{\theta}}}

\newcommand{\egcite}[1]{[e.g.,~\citealp{#1}]}
\usepackage[usenames, dvipsnames]{xcolor}

\usepackage{graphicx}
\usepackage{caption,subcaption}
\usepackage{mathtools}
\usepackage{amsfonts}
\usepackage[inline]{enumitem}  % inline enumeration
\usepackage{hyperref}
\usepackage[noabbrev]{cleveref}  % include the word "Figure" automatically
\usepackage{authblk}            % author affiliation etc
\hypersetup{
colorlinks=true,
citecolor=red
}

\usepackage{multicol}
\usepackage{tikz}
\usetikzlibrary{matrix,chains,positioning,decorations.pathreplacing,arrows,calc}
\usepackage{smartdiagram}
\usepackage[numbers]{natbib}  % plainnat bibliography style
\graphicspath{{pdf/}}

\begin{document}\sloppy

\title{Reduced order modeling of subsurface multiphase flow models using deep residual recurrent neural networks}

\author[]{J.Nagoor Kani}
\author[]{Ahmed H. Elsheikh}
%% \affil[]{School of Energy, Geoscience, Infrastructure and Society}
\affil[]{Heriot-Watt University, Edinburgh, UK}
\affil[]{\texttt{\{nj7,a.elsheikh\}@hw.ac.uk}}

\maketitle

\begin{abstract}
%{ \color{red}
We present a reduced order modeling (ROM) technique for subsurface multi-phase flow problems building on the recently introduced deep residual recurrent neural network (DR-RNN)~\citep{2017nagoor}. \mbox{DR-RNN} is a physics aware recurrent neural network for modeling the evolution of dynamical systems. The \mbox{DR-RNN} architecture is inspired by iterative update techniques of line search methods where a fixed number of layers are stacked together to minimize the residual (or reduced residual) of the physical model under consideration. In this manuscript, we combine \mbox{DR-RNN} with proper orthogonal decomposition (POD) and discrete empirical interpolation method (DEIM) to reduce the computational complexity associated with high-fidelity numerical simulations. In the presented formulation, POD is used to construct an optimal set of reduced basis functions and DEIM is employed to evaluate the nonlinear terms independent of the full-order model size.

We demonstrate the proposed reduced model on two uncertainty quantification test cases using Monte-Carlo simulation of subsurface flow with random permeability field. The obtained results demonstrate that \mbox{DR-RNN} combined with POD-DEIM provides an accurate and stable reduced model with a fixed computational budget that is much less than the computational cost of standard POD-Galerkin reduced model combined with DEIM for nonlinear dynamical systems.
\end{abstract}
%\end{frontmatter}

\section{Introduction}
\label{introduction}
Simulation of multi-phase flow in a subsurface porous media is an essential task for a number of engineering applications including ground water management, contaminant transport, and effective extraction of hydrocarbon resources~\citep{Petvipusit2014, Elsheikh2013-wrr}. The physics governing subsurface flow simulations are mainly modeled by a system of coupled nonlinear partial differential equations (PDEs) parametrized by subsurface properties (e.g. porosity and permeability)~\citep{aarnes2007introduction}.
In realistic settings, subsurface models are computationally expensive (i.e. large number of grid block is needed) as the subsurface properties are heterogeneous and the solution exhibit multiscale features~\citep{Elsheikh2012, Petvipusit2014}.

Moreover, these subsurface properties are only known at a sparse set of points (i.e. well locations) and the grid properties are populated stochastically over the entire domain~\citep{ibrahima2016,Elsheikh2012,Elsheikh2013-wrr}. Monte-Carlo methods are usually employed to propagate the uncertainties in the subsurface properties to the flow response. Monte-Carlo methods are computationally very expensive since a large number of forward simulations are necessary to estimate the statistics of the engineering quantities of interest~\citep{Petvipusit2014,Elsheikh2013-wrr,ibrahima2016}. Likewise, Bayesian inference tasks require a very large number of forward simulations to sharpen our knowledge about the unknown model parameters by utilizing field observation data~\citep{Elsheikh2012,Elsheikh2013-wrr}. For example, Markov-chain Monte-Carlo (MCMC) method (and its variants) requires a large number (in millions) of reservoir simulations to reach convergence and to avoid biased posterior estimates of the model parameters.

Surrogate models can be used to overcome the computational burden of multi-query tasks (e.g. uncertainty quantification, model based optimization) governed by large scale PDEs~\citep{podwilcox, koziel2013surrogate, he2013, Elsheikh2014, josset2015b, Hamid2015}. Surrogate models are computationally efficient mathematical models that can effectively approximate the main characteristics of the full-order model (full model)~\citep{podwilcox}. A number of surrogate modeling techniques have been developed and could be broadly classified into three classes: simplified physics based models~\citep{durlofsky2012uncertainty, josset2015b}, data-fit black-box models~\citep{podwilcox, li2017application, yeten2005comparison}, and projection based reduced order models commonly referred to as reduced model~\citep{podberkooz,podlassi,antoulas2001survey, Fang2013}. Physics based surrogate models are derived from high-fidelity models using approaches such as simplifying physics assumptions, using coarse grids, and/or upscaling of the model parameters~\citep{durlofsky2012uncertainty, podwilcox,he2013, Masoud2013}.
Data-fit models are generated using the detailed simulation data to regress the relation between the input and the corresponding output of interest~\citep{podwilcox,yeten2005comparison, abdi2018prediction, davidwood2018prediction}.
For a complete review of various surrogate modeling techniques, we refer the readers to the following papers by~\citet{asher2015review, podwilcox, koziel2013surrogate, razavi2012review}.

In projection based reduced order models (utilized in this paper), the governing equations of the full model are projected into a low-dimensional subspace spanned by a small set of basis functions via Galerkin projection~\citep{podlassi,antoulas2001survey}. Projection based ROMs relies on the assumption that most of the information and characteristics of the full model state variables can be efficiently represented by linear combinations of only a small number of basis functions. This assumption enables reduced model to accurately capture the input-output relationship of the full model with a significantly lower number of unknowns~\citep{podwilcox,podlassi,antoulas2001survey}. Projection based reduced order models are broadly categorized into: system based methods and snapshot based methods.
% based on the technique employed to compute the basis functions~\citep{antoulas2001survey,he2013}.
System based methods like balanced truncation realization methods~\citep{gugercin2004survey}, and Krylov subspace methods~\citep{freund2003model} use the characteristics of the full model and have been developed mainly for linear time-invariant problems, although much progress has been done on extensions of these methods to nonlinear problems~\citep{lall2002subspace}. Snapshot based methods such as reduced basis methods~\citep{rozza2007reduced}, proper orthogonal decomposition (POD)~\citep{podsiri,podberkooz} derive the projection bases from a set of full model solutions (the snapshots).

In this work, we employ POD based reduced model to accelerate Monte-Carlo simulation of subsurface flow models. The basis functions obtained from the POD is optimal in the sense that, for the same number of basis functions, no other bases can represent the given snapshot set with lower least--squares error than the POD bases~\citep{podlassi,podsiri} (see section~\ref{reducedordermodel} for further details).
\citet{lumley1967structure} was the first to apply POD techniques in fluid flow simulations. Since then, POD procedures has successfully been applied in a number of application areas~\egcite{podsiri, zheng2002low, cao2006reduced, bui2004aerodynamic, meyer2003efficient, astrid2004reduction, jin2018reduced}.

In the context of fluid flow in porous media,~\citet{vermeulen2004reduced} introduced POD in the confined, groundwater flow problems (linear subsurface flow model).~\citet{vermeulen2006model} applied POD in gradient-based optimization problem involving groundwater flow model.~\citet{mcphee2008groundwater} employed POD to enhance the groundwater management optimization problem.~\citet{siade2010snapshot} introduced a new methodology for the optimal selection of snapshots in such a way that the resulting POD basis functions account for the maximal variance of the full model solution.
Within the context of oil reservoir simulation,~\citet{heijn2003generation} and~\citet{van2006reduced} applied POD to accelerate the %the
optimization of a waterflood process.
\citet{cardoso2009development} incorporated a new snapshot clustering procedure to enhance the standard POD for oil–water subsurface flow problems.

In the context of Monte-Carlo simulations applied to stochastic subsurface flow problems, POD based ROMs were mainly employed only when the governing equation was linear (or nearly linear)~\citep{cardoso2010linearized, pasetto2011pod, pasetto2013reduced, boyce2014parameter}.~\citet{pasetto2011pod} employed POD based reduced model to construct MC realizations of two dimensional steady state confined groundwater flow subject to a spatially distributed random recharge.
\citet{pasetto2013reduced} applied POD to accelerate the MC simulations of transient confined groundwater flow models with stochastic hydraulic conductivity.~\citet{bau2012planning} derived a set of POD ROMs for each MC realization of hydraulic conductivity to solve a stochastic, multi-objective, confined groundwater management problem.~\citet{boyce2014parameter} applied a single parameter-independent POD reduced model to the deterministic inverse problem and the Bayesian inverse problem involving linear groundwater flow model.
In addition to the limitation of using only linear flow models, the UQ tasks in the aforementioned literature involve only low dimensional uncertain parameters.

Within the context of nonlinear subsurface flow problems, the target application of POD was mainly hydrocarbon production optimization, where POD ROMs were used mainly to optimize well control parameters (e.g., bottomhole pressure)~\citep{cardoso2010linearized, he2011enhanced, trehan2016trajectory, rousset2014reduced, jansen2017use}. Recently,~\citet{jansen2017use} has done an extensive review on the use of reduced-order models in well control optimization. For the well control applications, POD achieved reasonable levels of accuracy only when the well controls in test runs were relatively close to those used in training runs. In the case where the test controls substantially differ from those used in the initial training runs, additional computational steps were needed. For example refitting the POD basis functions was performed in~\citep{trehan2016trajectory}, which impose some additional computational overhead.
Although POD combined with Galerkin projection has been applied more frequently to nonlinear flow problems~\citep{bui2004aerodynamic, podberkooz, rousset2014reduced}, the effectiveness of POD-Galerkin based model in handling nonlinear systems is limited mainly by two factors. The first factor is related to the treatment of the nonlinear terms in the POD-Galerkin reduced model~\citep{deim,tpwl,cardoso2010linearized} and the second factor is related to maintaining the overall stability of the resulting reduced model~\citep{cardoso2010linearized, he2010, he2013, 2007goal,wang2012}.

In relation to computing reduced non-polynomial nonlinear functions, POD based ROMs is usually dependent on the full model state variables and henceforth, the computational cost of evaluating the reduced model is still a function of full model dimension. Several techniques have been developed to reduce the computational cost of evaluating the nonlinear term in POD ROMs including trajectory piecewise linearization (TPWL)~\citep{tpwl}, gappy POD technique~\citep{willcox2006unsteady}, missing point estimation (MPE)~\citep{barrault2004empirical}, best point interpolation method~\citep{nguyen2008best}, and discrete empirical interpolation method (DEIM)~\citep{barrault2004empirical,deim}. Among these techniques, TPWL and DEIM are widely used for efficient treatment of nonlinearities in multi-phase flow reservoir simulations~\citep{gasemi2015,he2010,he2013}.

In TPWL method~\citep{tpwl}, the nonlinear function is first approximated by a piecewise linear function obtained by linearizing the full-order model at a predetermined set of points in the time and the parameter space. Then the nonlinear full model is replaced by an adequately weighted sum of the selected linearized systems~\citep{tpwl}. Finally, the reduced model can be obtained by projecting the resultant linearized full-order system using standard techniques like POD~\citep{tpwl}. The TPWL method was first introduced in~\citep{tpwl} for modeling nonlinear circuits and micromachined devices.
In the context of subsurface flow problems, TPWL procedures were applied in~\citep{cardoso2010linearized, he2011enhanced, trehan2016trajectory, rousset2014reduced} to accelerate the solution of production optimization problems.

In DEIM, the nonlinear term in the full model is approximated by a linear combination of a set of basis vectors~\citep{deim}. The coefficients of expansion are determined by evaluating the nonlinear term only at a small number of selected interpolation points~\citep{deim}.
DEIM was developed in~\citep{deim} for model reduction of general nonlinear system of ordinary differential equations (ODEs) and have been used in several areas~\citep{chaturantabut2012state, xiao2014non, buffoni2010projection}. Within the context of subsurface flow problems,~\citet{chaturantabut2011application} applied DEIM for model reduction of viscous fingering problems of an incompressible fluid through a two dimensional homogeneous porous medium.
\citet{alghareeb2013optimum} combined DEIM with POD procedures and the resultant reduced model was applied in waterflood optimization problem.
Recently,~\citet{gasemi2015} applied POD with DEIM to an optimal control problem governed by two-phase flow in a porous media. Next,~\citet{gasemi2015} used machine learning technique to construct a number of POD-DEIM local reduced-order models. In that work, machine learning technique was used to construct a number of POD-DEIM local reduced-order models and then a specific local reduced-order model was selected with respect to the current state of the dynamical system during the gradient based optimization task. Similarly,~\citet{yoon2014development} used multiple local DEIM approximations in POD reduced model framework to reduce the computational costs of high-fidelity reservoir simulations.

The overall convergence and stability is another issue that limits the applicability of POD-Galerkin based ROMs. POD-Galerkin projection methods manage to decrease the computational complexity by orders of magnitude as a result of state variable's dimension reduction. However, this reduction goes hand in hand with a loss in accuracy. Moreover, slow convergence and in some cases model instabilities~\citep{wang2012, he2010,2007goal} are observed as the errors in the reduced state variables are propagated in time. More specifically, the performance of POD-Galerkin ROMs is directly influenced by the number of POD basis used in the POD-Galerkin projection. However, in many applications involving nonlinear conservation laws (e.g. high Reynold number fluid flow), POD-Galerkin reduced order models have shown poor performance even after retaining a sufficient number of POD basis~\citep{wang2012, podsiri,podberkooz}.

Several stabilization techniques have been proposed in the recent literature to build a stabilized POD based reduced models. A notable stabilization technique relies on closing the POD reduced model using a set of closure models similar to those adopted in turbulence modeling~\citep{podberkooz, wang2012}. The objective of applying closure models within POD based reduced model is to include the effects of the discarded POD basis functions in the extracted reduced model~\citep{podberkooz, wang2012}.
\citet{wang2012} showed that POD-Galerkin reduced model yielded inaccurate and physically implausible results when applied to the numerical simulation of a 3D turbulent flow past a cylinder at Reynold number of $1000$.~\citet{wang2012} addressed the aforementioned accuracy and stability issues of POD reduced model by various closure models, where artificial viscosity was added to the real viscosity parameter to stabilize the POD based reduced model.

Another major approach to enhance the stability of POD-Galerkin reduced model is to compute a new set of optimal basis or to improve the POD basis vectors by solving a constrained optimization problem.
\citet{2007goal} determined a new set of optimal basis vectors by formulating an optimization problem constrained by the equations of the resultant reduced model and demonstrated the stability of the proposed approach on linear dynamical systems. We note that POD-Galerkin reduced model orthogonally projects the nonlinear residual into the subspace spanned by the POD basis vectors. Unlike POD-Galerkin reduced model, Petrov-Galerkin projection scheme design a different set of orthonormal basis called left reduced order basis into which the nonlinear residual is projected.~\citet{carlberg} formulated stable Petrov-Galerkin reduced model in which the left reduced order basis vectors were computed from an optimization problem at every iteration of the Gauss Newton method.~\citet{he2010} observed that poor spectral properties of the reduced Jacobian matrix could cause numerical instabilities in POD-Galerkin TPWL reduced model. Hence,~\citet{he2010} improved the stability of the POD based reduced model by determining the optimal dimension of the reduced model through an extensive search over a range of integer numbers. We note that all the above mentioned optimization procedures involve computationally expensive procedures to maintain stability and in many cases, the stability of the extracted reduced model is still not guaranteed~\citep{he2010, he2013}.

Recently, data-fit black-box models have been combined with POD~\citep{xiao2017non} to develop non-intrusive POD based ROMs, where the data-fit models are used to regress the relationship between the input parameter and the reduced representation of the full model state vector. Hence, non-intrusive ROMs do not require any knowledge of the full-order model and is mainly developed to circumvent the shortcomings in accessing the governing equations of the full model~\citep{xiao2017non}. However, it can also be used to address the stability and nonlinearity issues of POD based ROMs.
\citet{wang2017model} developed a non-intrusive POD reduced model using Recurrent Neural Network (RNN) as a data-fit model and presented two fluid dynamics test cases namely, flow past a cylinder
and a simplified wind driven ocean gyre.
RNN is a class of artificial neural network~\citep{pascanu2013difficulty,mikolov2014learning} which has at least one feedback connection in addition to the feedforward connections~\citep{pascanu2013construct, pascanu2013difficulty,irsoy2014opinion}. In the context of data-fit models, RNN has been successfully applied to various sequence modeling tasks such as automatic speech recognition and system identification of time series data~\citep{hermans2013training,he2015deep,hinton2012deep,graves2013generating}. Additionally, RNN has been applied to emulate the evolution of nonlinear dynamical systems in a number of applications~\citep{zimmermann2012forecasting,bailordavidrecurrent} and henceforth has large potential in building reduced order models. However, the applicability of non-intrusive ROMs is severely undermined in many real-world problems, where increasing the dimensionality of the input parameter space increases the complexity and training time of the data-fit model.

In summary, among many surrogate modeling techniques, POD-Galerkin reduced model is a viable option for accelerating multi-query tasks like UQ. Generally, POD-Galerkin reduced model is well established for linear systems and for nonlinear systems with parametric dependence, POD could be either combined with TPWL or with DEIM for modeling subsurface flow systems~\citep{cardoso2010linearized, he2011enhanced, trehan2016trajectory, gasemi2015}. However, POD reduced model does not preserve the stability properties of the corresponding full order model and current state of the art POD stabilization techniques~\citep{wang2012, he2010, he2013} are not cost effective and ultimately do not guarantee stability of the extracted reduced order models.

In this paper, we use \mbox{DR-RNN}~\citep{2017nagoor} to alleviate the potential limitations of POD-Galerkin reduced models. More specifically, we combine~\mbox{DR-RNN} with POD-Galerkin and DEIM methods to derive an accurate and computationally effective reduced model for uncertainty quantification (UQ) tasks. The architecture of \mbox{DR-RNN} is inspired by the iterative line search methods where the parameters of the \mbox{DR-RNN} are optimized such that the residual of the numerically discretized PDEs is minimized~\citep{bertsekas1999nonlinear, tieleman2012lecture, 2017nagoor}. Unlike the standard RNN which is very generic, DR-RNN~\citep{2017nagoor} uses the residual of the discretized differential equation. In addition, the parameters of the DR-RNN are fitted such that the computed DR-RNN output optimally minimizes the residual of the targeted equation. In this context, DR-RNN is a physics aware RNN as it is tailored to leverage the physics embedded in the targeted dynamical system (i.e. residual of the equation or reduced residual in the current manuscript).

The resultant reduced model obtained from \mbox{DR-RNN} combined with POD-Galerkin and DEIM algorithm has a number of salient features. First, the dynamics of \mbox{DR-RNN} is explicit in time with superior convergence and stability properties for large time steps that violate the numerical stability conditions~\citep{2017nagoor, pletcher2012computational}. Second, as the dynamics modeled in \mbox{DR-RNN} are explicit in time, there is a reduction in the computational complexity of the extracted reduced model from $\mathcal{O}(r^3)$ corresponding to implicit POD-DEIM reduced order models, to $\mathcal{O}(r^2)$, where $r$ is the size of the reduced model. Third, \mbox{DR-RNN} requires only very few training samples (obtained by solving the full model) to optimize the parameters of the DR-RNN as it accounts for the physics of the full model within the RNN architecture (via the reduced residual). This is a major advantage when compared to pure data-driven algorithms (e.g. standard RNN architectures). Moreover, \mbox{DR-RNN} can effectively emulate the parameterized nonlinear dynamical system with a significantly lower number of parameters in comparison to standard RNN architectures~\citep{2017nagoor}.

In this work, we demonstrate the superior properties of \mbox{DR-RNN} in accelerating UQ tasks for subsurface reservoir models using Monte-Carlo method. As far as we are aware, the use of a single parameter-independent POD-Galerkin reduced model in Monte-Carlo method involving nonlinear subsurface flow with high dimensional stochastic permeability field has not been previously explored. The reason is that the resultant reduced model might require significantly more basis functions to reconstruct stable solutions~\citep{cardoso2010linearized, he2011enhanced, boyce2014parameter, gasemi2015}. However, only a single set of small number of POD basis functions would be sufficient to reconstruct the solution with reasonable accuracy using least--squares (see section~\ref{leastsquaresection} for more details). Hence, the aim of this paper is to illustrate how \mbox{DR-RNN} could be used to reconstruct stable solutions emulating the full model dynamics using only a small set of POD basis functions. The proposed \mbox{DR-RNN} technique is validated on two forward uncertainty quantification problems involving two-phase flow in subsurface porous media. The two flow problems are commonly known within the reservoir simulation community as the quarter five spot problem and the uniform flow problem~\citep{aarnes2007introduction}. In these two numerical examples, the permeability field is modeled as log-normal distribution. The obtained results demonstrate that \mbox{DR-RNN} combined with POD-DEIM provides an accurate and stable reduced order model with a drastic reduction in the computational cost. The reason for selecting simplified flow problems is to illustrate the potential benefit of DR-RNN to formulate an accurate and computationally effective POD-DEIM reduced model for flow problems where the standard POD-Galerkin reduced models are inaccurate and possibly unstable. We also note that DR-RNN architecture is generic and could be used to emulate any well-posed nonlinear dynamical system~\citep{2017nagoor} including subsurface flow problems while accounting for capillary pressure effects, gravity effects and compressibility.

The outline of the rest of this manuscript is as follows: In section~\ref{problemsetup}, we present the formulation of multi-phase flow problem in a porous media. In section~\ref{reducedordermodel}, we introduce POD-Galerkin method for model reduction followed by a discussion of DEIM for handling nonlinear systems. In Section~\ref{drrnnsec}, we describe the architecture of \mbox{DR-RNN} and in section~\ref{numericalexamples}, we evaluate the reduced model derived by combining \mbox{DR-RNN} with POD-DEIM on two uncertainty quantification test cases. Finally, in Section~\ref{sec:conclusions}, we present the conclusions of this manuscript.

\section{Problem Formulation}
\label{problemsetup}
The equations governing two-phase flow of a wetting phase (water) and non-wetting phase (e.g. oil) in a porous media are the conservation of mass (continuity) equation and Darcy's law for each phase ~\citep{aarnes2007introduction, he2013, chen2006computational, bastian}. The continuity equation for each phase $\alpha$ takes the form
%%%%%%%%%%%%%%%%%%%%%%%%%%%%%%%%%
\begin{equation}
\dfrac{\partial(\phi \rho_{\alpha} s_{\alpha}) }{ \partial t} - \nabla \cdot (\rho_{\alpha} \lambda_{\alpha} \mK~(\nabla p_{\alpha} - \rho_{\alpha} g \nabla h) ) + q_{\alpha} = 0
\label{continuity_eq}
\end{equation}
%%%%%%%%%%%%%%%%%%%%%%%%%%%%%%%%%%
where the subscript $\alpha=w$ denotes the water phase, the subscript $\alpha=o$ denotes the oil phase, $\mK$ is the absolute permeability tensor, $\lambda_{\alpha} = k_{r\alpha}/\mu_{\alpha}$ is the phase mobility, with $k_{r\alpha}$ the relative permeability to phase $\alpha$ and $\mu_{\alpha}$ the viscosity of phase $\alpha$, $p_{\alpha}$ is the phase pressure, $\rho_{\alpha}$ is the density of phase $\alpha$, $g$ is the gravitational acceleration, $h$ is the depth, $\phi$ is the porosity, $s_{\alpha}$ is the saturation of the phase $\alpha$ and $q_{\alpha}$ is the phase source and sink terms~\citep{aarnes2007introduction, chen2006computational}. Further, the phase saturations are constrained by $s_w + s_o = 1$, since the oil and the water jointly fill the void space~\citep{aarnes2007introduction, he2013}.

The phase velocities are modeled by the multiphase Darcy's law to relate the phase velocities to the phase pressures and takes the form
%%%%%%%%%%%%%%%%%%%%%%%%%%%%%%%%%%%%%%%
\begin{equation}
\vv_{\alpha} = -\mK \lambda_{\alpha}~\nabla(p_{\alpha} - \rho_{\alpha} g h)
\label{darcy_eq}
\end{equation}
%%%%%%%%%%%%%%%%%%%%%%%%%%%%%%%%%%%%%%
where $\vv_{\alpha}$ is the phase velocity. The phase relative permeabilities $k_{r\alpha}$ and the capillary pressure ($ p_{cow} = p_o - p_w$) are usually modeled as functions of the phase saturations~\citep{aarnes2007introduction}. Neglecting the capillary pressure, the compressibility effects, the gravitational effects, and assuming the density ratio to be equal to one, the continuity equations (Eq.~\eqref{continuity_eq}) can be combined with the Darcy's law (Eq.~\eqref{darcy_eq}) to derive a global pressure equation and the saturation equation for water phase~\citep{aarnes2007introduction,he2013, bastian}. The simplified global pressure equation takes the form
%%%%%%%%%%%%%%%%%%%%%%%%%%%%%%%%%%%%%%%%
\begin{equation}
\nabla \cdot \mK \lambda~\nabla p = q
\label{pressure_eq}
\end{equation}
%%%%%%%%%%%%%%%%%%%%%%%%%%%%%%%%%%%%%%%%
where $p=p_o=p_w$ is the global pressure, $\lambda = \lambda_w + \lambda_o$ is the total mobility, $q=q_w+q_o$ is the source and sink term. The saturation equation for the water phase takes the following form
%%%%%%%%%%%%%%%%%%%%%%%%%%%%%%%%%%%%%%%%%%%%%%%%%%%%%%%%%%%%%%%%%%%%%%%%%%%%%%%%%%
\begin{equation}
\phi~\dfrac{\partial s}{\partial t} + \vv \cdot \nabla f_w = \frac{q_w}{\rho_w}
\label{saturation_eq}
\end{equation}
%%%%%%%%%%%%%%%%%%%%%%%%%%%%%%%%%%%%%%%%%%%%%%%%%%%%%%%%%%%%%%%%%%%%%%%%%%%%%%%%%%
where $f_w = {\lambda_w}/(\lambda_w + \lambda_o)$ is a function of saturation termed as the fractional flow function for the water phase, $\vv=-\mK\lambda~\nabla p$ is the total velocity vector and $s=s_w$ is the water saturation~\citep{aarnes2007introduction, chen2006computational}. In the rest of the paper, we write the water phase saturation as $s=s_w$ for simplicity. The coupled equations Eq.~\eqref{pressure_eq} and Eq.~\eqref{saturation_eq} could then be solved for the evolution of the saturation by providing the appropriate initial and boundary conditions. Equation~\eqref{pressure_eq} and Eq.~\eqref{saturation_eq} are continuous (in space and time) form of the full model.

The discrete form of the full model is obtained by dividing the problem domain into $n$ grid blocks and then applying the finite volume method to discretize the spatial derivatives of Eq.~\eqref{pressure_eq} and Eq.~\eqref{saturation_eq}. The discretized pressure equation takes the form
%%%%%%%%%%%%%%%%%%%%%%%%%%%%%%%%%%%%%%%
\begin{equation}
\mA~\vy_p = \vb
\label{FOM1}
\end{equation}
%%%%%%%%%%%%%%%%%%%%%%%%%%%%%%%%%%%%%%
where $\mA \in \mathbb{R}^{n \times n}$, $\vb \in \mathbb{R}^n$, and $\vy_p \in \mathbb{R}^n$ is the pressure vector in which each component $\mathit{y_{p_i}}$ of $\vy_p$ represent the pressure value at the $i$th grid block. Similarly, the spatially discretized saturation equation takes the form
%%%%%%%%%%%%%%%%%%%%%%%%%%%%%%%%%%%%%%%%%%%%%
\begin{equation}
\dfrac{d \vy_s}{dt} + \mB(\vv)~\vf_w(\vy_s) = \vd
\label{FOM2}
\end{equation}
%%%%%%%%%%%%%%%%%%%%%%%%%%%%%%%%%%%%%%%%%%%%
where $\mB \in \mathbb{R}^{n \times n}$, $\vd \in \mathbb{R}^n$, $\vv$ is the total velocity vector, and $\vy_s \in \mathbb{R}^n$ is the saturation vector in which each component $\mathit{y_{s_i}}$ of $\vy_s$ is the saturation value at the $i$th grid block.

Eq.~\eqref{FOM1} and Eq.~\eqref{FOM2} are the discrete form of the full model for multi-phase flow problem under consideration. These two equations exhibit two way coupling from the dependence of the matrix $\mA$ on the mobilities $\lambda(\vy_s(t))$ in the pressure full model (Eq.~\eqref{FOM1}) and from the dependence of the matrix $\mB$ on the velocity vector $\vv(\vy_p)$ in the saturation full model (Eq.~\eqref{FOM2}). In this paper, we adopt an implicit sequential splitting method to solve the full model (Eq.~\eqref{FOM1} and Eq.~\eqref{FOM2}). In this method, the saturation vector $\vy_s(t)$ from the present time step is used to assemble the matrix $\mA$ in Eq.~\eqref{FOM1} and then the pressure full model (Eq.~\eqref{FOM1}) is solved for the pressure vector $\vy_p$. Following that, the velocity vector $\vv$ (computed from the pressure vector $\vy_p$) is used to assemble the matrix $\mB$ in Eq.~\eqref{FOM2} and then the saturation full model (Eq.~\eqref{FOM2}) is solved implicitly in time for the saturation at the next time step. In the following section, we formulate a Galerkin projection based reduced model to reduce the computational effort for multi-query tasks (e.g. uncertainty quantification) involving repeated solutions of Eq.~\eqref{FOM1} and Eq.~\eqref{FOM2}, when $n$ (the number of grid block) is large~\citep{deim,gasemi2015}.

\section{Reduced Order Model Formulation}
\label{reducedordermodel}
In this section, we formulate the POD-Galerkin reduced model (POD reduced model) and POD-DEIM reduced model where POD-Galerkin is combined with DEIM for handling the nonlinear terms. Both methods are introduced to reduce the computational effort associated with solving the full model (Eq.~\eqref{FOM1} and Eq.~\eqref{FOM2}).
%%%%%%%%%%%%%%%%%%%%%%%%%%%%%%%%%%%%%%
\subsection{POD basis}
%%%%%%%%%%%%%%%%%%%%%%%%%%%%%%%%%%%%%%
As stated in section~\ref{introduction}, POD based reduced model is a projection based reduced order model in which the governing equations are projected onto an optimal low-dimensional subspace $\mathcal{U}$ spanned by a small set of $r$ basis vectors. Galerkin projection reduced model is based on the assumption that most of the system information and characteristics can be efficiently represented by linear combinations of only a small number of basis vectors~\citep{tpwl}.

The optimal basis vectors $\lbrace \vu_i \rbrace_{i=1}^{r}$ in POD are computed by singular value decomposition (SVD) of the solution snapshot matrix $\mX$. The solution snapshot matrix $\mX$ is obtained from a set of solution vectors of size $n_s$ obtained by solving the full model at selected points in the input parameter space. The SVD of $\mX$ is expressed as
\begin{equation}
\mX = \mU~\Sigma~\mW
\label{SVD_eq}
\end{equation}
where, $\mX \in \mathbb{R}^{n \times n_s}$, $\mU=[\vu_1~\vu_2~\vu_3~\cdots~\vu_n] \in \mathbb{R}^{n \times n}$ is the left singular matrix and $\Sigma=\text{diag}(\sigma_1>\sigma_2>\sigma_3>\cdots~\sigma_{ns} \geq 0)$ is the diagonal matrix containing the singular values $\sigma_i$ of the snapshot matrix $\mX$ in descending order.
The dominant left singular vectors $\lbrace \vu_i \rbrace_{i=1}^{r}$ corresponding to the first $r$ largest singular values represents the basis vectors to span the optimal subspace $\mathcal{U}$ of POD based reduced model. Thus, the first step in deriving the POD based reduced model is to express the state vector $\vy$ of the full-order model by a linear combination of $r$ basis vectors as following
%%%%%%%%%%%%%%%%%%%%%%%%%%%%%%%
\begin{equation}
\label{svdfit_eq}
\vy \approx \mU^r~\tilde{\vy}
\end{equation}
%%%%%%%%%%%%%%%%%%%%%%%%%%%%%%%
where $\tilde{\vy} \in \mathbb{R}^r$ is the reduced state vector representation of full dimensional state vector $\vy$, and $\mU^r=[\vu_1~\cdots~\vu_r] \in \mathbb{R}^{n \times r}$ is the matrix that contains $r$ orthonormal basis vectors in its columns.

By following this step, for example, the optimal basis vectors for the saturation state vector $\vy_s$ are obtained from the SVD of the saturation snapshot matrix $\mX_{s}=\left((\vy_{s_1}~\ldots~\vy_{s_T})^1~\ldots~(\vy_{s_1}~\ldots~\vy_{s_T})^L\right)$, where $T$ is the number of time steps and $L$ is the number of samples of input parameter used to build the snapshot matrix. The SVD of $\mX_{s}$ is expressed as
%%%%%%%%%%%%%%%%%%%%%%%%%%%%%%%%%%%%%%%%%
\begin{equation}
\mX_{s} = \mU_{s}~\Sigma_{s}~\mW_{s}
\label{SVD1_eq}
\end{equation}
%%%%%%%%%%%%%%%%%%%%%%%%%%%%%%%%%%%%%%%%
where $\mU_{s} \in \mathbb{R}^{n \times n}$ is the left singular matrix, $\Sigma_{s}$ is the diagonal matrix containing the singular values of the snapshot matrix $\mX_{s}$ in descending order. The saturation state vector $\vy_s$ is optimally expressed as
%%%%%%%%%%%%%%%%%%%%%%%%%%%%%%%
\begin{equation}
\vy_s \approx \mU_{s}^r~\tilde{\vy}_s
\label{svdfit12_eq}
\end{equation}
%\label{svdfit1_eq}
%%%%%%%%%%%%%%%%%%%%%%%%%%%%%%%
where $\tilde{\vy}_s \in \mathbb{R}^r$ is the reduced state vector representation of $\vy_s$, $\mU_{s}^r \in \mathcal{R}^{n \times r}$ is the matrix that contains $r$ orthonormal basis vectors in its columns. Similarly, we can represent the pressure state vector $\vy_p$ from its reduced state vector representation $\tilde{\vy}_p$ using optimal basis matrix $\mU_p$ obtained from the SVD of the pressure snapshot matrix $\mX_p$.

%%%%%%%%%%%%%%%%%%%%%%%%%%%%%%%%%%%%%%%%
%\subsection*{Best approximation}
\subsection{Least--squares approximation}
\label{leastsquaresection}
%%%%%%%%%%%%%%%%%%%%%%%%%%%%%%%%%%%%%%%%%%%
The capacity of a set of basis functions to represent a new solution vector could be tested using least--squares fitting~\citep{elden2007matrix, trefethennumerical}. For example, the least--squares solution for approximating a saturation state vector $\vy_s^* \in \mathbb{R}^n$ is defined as
\begin{equation}
\vy_s^{*} \approx \mU_s^r~\tilde{\vy}_s = \mU_s^r~({\mU_s^r}^{\top}~\vy_s)
\label{Least--squaresolution_eq}
\end{equation}
%%%%%%%%%%%%%%%%%%%%%%%%%%%%%%%%%%%%%%%%
The associated error termed as least--squares errors in approximating $\vy_s$ by $\vy_s^*$ using only $r$ basis vectors is given by
%%%%%%%%%%%%%%%%%%%%%%%%%%%%%%%%%%%%%%%%%%%%%%%%%%
\begin{equation}
\varepsilon_s = \Vert \vy_s - \vy_s^{*} \Vert_2
\label{leastsquareerror_eq}
\end{equation}
%%%%%%%%%%%%%%%%%%%%%%%%%%%%%%%%%%%%%%%%%%%%%%%%%%
The least--squares error $\varepsilon_s$ (Eq.~\eqref{leastsquareerror_eq}) is equivalent to the omitted energy $\Omega_s = \sum_{i=r+1}^{n} \sigma_{s_i}$~\citep{lucia2004reduced, podberkooz}. In practice, $r$ is commonly chosen as the smallest integer such that the relative omitted energy $\nu$ is less than a preset value (e.g. $0.01$), where the omitted energy is defined by the following equation
%%%%%%%%%%%%%%%%%%%%%%%%%%%%%%%%%%%%%%%%%%%%%%%%%%%%%%%%%%%%%
\begin{equation}
\nu = 1 - \frac{\sum_{i=r+1}^{n}\sigma_{s_i}}{\sum_{i=1}^{n}\sigma_{s_i}}
\label{omittedenergy_eq}
\end{equation}
%%%%%%%%%%%%%%%%%%%%%%%%%%%%%%%%%%%%%%%%%%%%%%%%%%%%%%%%%%%%%
Similar expressions mentioned in Eq.~\eqref{Least--squaresolution_eq}, Eq.~\eqref{leastsquareerror_eq}, and Eq.~\eqref{omittedenergy_eq} can be obtained for the pressure state vector as well.
We note that least--squares errors are not necessarily equivalent to the omitted energy for state vectors not included in the snapshot matrix or for the state vector solved at a new point in the input parameter space as these new vectors might not fall within the span of the snapshot matrix~\citep{podwilcox,lucia2004reduced}. The least--squares solution is the best approximation of the state variables in the sense that, for the chosen low dimensional subspace $\mathcal{U}$, no other low dimensional approximation can represent the given snapshot set with a lower least--squares error~\citep{podlassi,podsiri,podberkooz}. In this paper, we use the best approximation of the state variables to assess the quality of the approximation obtained from different reduced order models in the numerical examples presented in section~\ref{numericalexamples}.

%%%%%%%%%%%%%%%%%%%%%%%%%%%%%%%%%%%%%%%
\subsection{POD-Galerkin}
%%%%%%%%%%%%%%%%%%%%%%%%%%%%%%%%%%%%%%%%
Once the POD basis vectors are obtained, the reduced representation of the pressure vector $\vy_p$ is substituted into the pressure full model (Eq.~\eqref{FOM1}), followed by Galerkin projection of the pressure equation into the subspace spanned by $\mU^r_{p}$. The resulting POD based reduced model for the pressure equation then takes the following form
%%%%%%%%%%%%%%%%%%%%%%%%%%%%%%%%%%%%%%%%%%%
\begin{equation}
\tilde{\mA}~\tilde{\vy}_p = \tilde{\vb}
\label{ROM1_eq}
\end{equation}
%%%%%%%%%%%%%%%%%%%%%%%%%%%%%%%%%%%%%%%%%%
where $\tilde{\mA} = {\mU_{p}^r}^{\top}~\mA~\mU_{p}^r \in \mathbb{R}^{r \times r}$ and $\tilde{\vb}={\mU_{p}^r}^{\top}~\vb \in \mathbb{R}^r$.
Similarly, POD based reduced model for the saturation equation (Eq.~\eqref{FOM2}) takes the form
%%%%%%%%%%%%%%%%%%%%%%%%%%%%%%%%%%%%%%%%%%%%%%%%%%%%%%%%%%%%%%%%%%%%%%%%%%%%%%%%%%%%%%%%%
\begin{equation}
\dfrac{d \tilde{\vy}_s}{dt} + {\mU_{s}^r}^{\top}~\mB(\vv)~\vf_w(\mU_{s}^r~\tilde{\vy}_s) = \tilde{\vd},
\label{ROM2_eq}
\end{equation}
%%%%%%%%%%%%%%%%%%%%%%%%%%%%%%%%%%%%%%%%%%%%%%%%%%%%%%%%%%%%%%%%%%%%%%%%%%%%%%%%%%%%%%%%%
where $\tilde{\vd}={\mU_{s}^r}^{\top}~\vd$ and $\tilde{\vd} \in \mathbb{R}^r$.
%\end{document}

The POD based reduced model formulated by Eq.~\eqref{ROM1_eq} and Eq.~\eqref{ROM2_eq} is of the reduced dimension $r$. However, the nonlinear function $\vf_w$ in Eq.~\eqref{ROM2_eq} is still of the order of full dimension $n$. Moreover, the reduced Jacobian matrix $ \tilde{\mJ} = \tilde{\mI} - {\mU_s^r}^{\top}\mB~\mJ_f(\vf_w(\mU_s^r~\tilde{\vy}_s))\mU_s^r~\in \mathbb{R}^{r \times r}$ needed for Newton like iterations to solve this nonlinear equation is also of order $n$~\citep{deim} as it relies on evaluating the full order nonlinear function $\vf_w$. Therefore, for problems with general nonlinear functions involved in POD based reduced model, the computational cost of solving the reduced system is still a function of the full system dimension $n$.
%\end{document}
%%%%%%%%%%%%%%%%%%%%%%%%%%%%%%%%%%%%%%%
\subsection{DEIM}
%%%%%%%%%%%%%%%%%%%%%%%%%%%%%%%%%%%%%%%
 Discrete Empirical Interpolation Method (DEIM) was introduced in~\citep{deim} to approximate the nonlinear terms in POD based reduced model using a limited number of points that are independent of the full system dimension $n$. Similar to POD, the first step of DEIM is to approximate the nonlinear function $\vf_w$ in Eq.~\eqref{ROM2_eq} using a separate set of basis vectors $\mV^m=[\vv_1~\vv_2~\vv_3~\ldots~\vv_m]$ as
%%%%%%%%%%%%%%%%%%%%%%%%%
\begin{equation}
\label{svdfitfw_eq}
\vf_w = \mV^m~\tilde{\vf}
\end{equation}
%%%%%%%%%%%%%%%%%%%%%%%%%
where $\tilde{\vf}$ is the coefficient of expansion of the nonlinear function $\vf_w$ in the reduced subspace spanned by $\lbrace \vv_i \rbrace_{i=1}^m$, $\mV^m \in \mathbb{R}^{n \times m}$ is the matrix containing the first $m$ columns of the left singular matrix $\mV \in \mathbb{R}^{n \times n}$ obtained from the SVD of the the snapshot matrix $\mX_{f}$ of the nonlinear function $\vf_w$. We note that no additional computational costs are associated with collecting the snapshot matrix of the nonlinear terms $\mX_f$ as it is already evaluated during the computation of the state snapshot vectors. The nonlinear term in Eq.~\eqref{ROM2_eq} can then be expressed as
%%%%%%%%%%%%%%%%%%%%%%%%%%%%%%%%%%%
\begin{equation}
{\mU_{s}^r}^{\top}~\mB~\vf_w = ({\mU_{s}^r}^{\top}~\mB~\mV^m)~\tilde{\vf} =
({\mU_{s}^r}^{\top}~\mB~\mV^m)~\cdot~({\mV^m}^\top~\vf_w)
\label{nonlinear_eq}
\end{equation}
%%%%%%%%%%%%%%%%%%%%%%%%%%%%%%%%%%%%
The matrix factor $({\mU_{s}^r}^{\top}~\mB~\mV^m) \in \mathbb{R}^{r \times m} $ in Eq.~\eqref{nonlinear_eq} is precomputed before solving Eq.~\eqref{ROM2_eq}. The overdetermined system $\tilde{\vf} = {\mV^m}^\top~\vf_w$ is approximated using the DEIM algorithm introduced in~\citep{deim} by first computing a matrix $\mP \in \mathbb{R}^{n \times m}$ that selects $m$ rows of the matrix $\mV^m$ to obtain $\tilde{\vf}$ as following
%%%%%%%%%%%%%%%%%%%%%%%%%%%%%%%%%%%
\begin{equation}
\mP^{\top}~\vf_w = \mP^{\top}~\mV^m~\tilde{\vf} \rightarrow \tilde{\vf} = (\mP^{\top}~\mV^m)^{-1}~\mP^{\top}~\vf_w
\label{deimfw_eq}
\end{equation}
%%%%%%%%%%%%%%%%%%%%%%%%%%%%%%%%%%%%
Using this expression of $\tilde{\vf}$ to approximate the nonlinear function in Eq.~\eqref{nonlinear_eq}, we obtain a nonlinear term that is independent of $n$ that takes the form
%%%%%%%%%%%%%%%%%%%%%%%%%%%%%%%%%%%%%
\begin{equation}
{\mU_{s}^r}^{\top}~\mB~\vf_w \approx \mD~\vf_w(\mP^{\top}~\mU_{s}^r~\tilde{\vy}_s)
\end{equation}
%%%%%%%%%%%%%%%%%%%%%%%%%%%%%%%%%%%%%
where the matrix $\mD = {\mU_{s}^r}^{\top}~\mB~\mV^m~(\mP^{\top}~\mV^m)^{-1} \in \mathbb{R}^{r \times m}$ termed as the DEIM matrix. Similarly, the Jacobian of the nonlinear term in Eq.~\eqref{ROM2_eq} is approximated using DEIM as following
%%%%%%%%%%%%%%%%%%%%%%%%%%%%%%%%
\begin{equation}
\tilde{\mJ} = \tilde{\mI} - ({\mU_{s}^r}^{\top}\mB\mV^m(\mP^{\top}~\mV^m)^{-1})~\hat{\mJ}_f(\vf_w(\mP^{\top}~\mU_{s}^r~\tilde{\vy}_s))~(\mP^{\top}
\mU_{s}^r)
\end{equation}
%%%%%%%%%%%%%%%%%%%%%%%%%%%%%%%%
where $\hat{\mJ}_f(\vf_w(\mP^{\top}~\mU_{s}^r~\tilde{\vy}_s)) \in \mathbb{R}^{m \times m}$ is the Jacobian matrix computed using the $m$ components of $\vf_w$ evaluated by the DEIM algorithm~\citep{deim, tpwl, 2017nagoor}. Finally, the POD-DEIM based reduced model takes the form
%%%%%%%%%%%%%%%%%%%%%%%%%%%%%%%%%%%%%%%%%%%%%%%%%%%%%%%%%%%%%%%%%%%%%%%%%%%%%%%%%%%%%%%%%
\begin{equation}
\dfrac{d \tilde{\vy}_s}{dt} + \mD~\vf_w(\mP^{\top}~\mU_{s}^r~\tilde{\vy}_s) = \tilde{\vd}
\label{ROMDEIM_eq}
\end{equation}
%%%%%%%%%%%%%%%%%%%%%%%%%%%%%%%%%%%%%%%%%%%%%%%%%%%%%%%%%%%%%%%%%%%%%%%%%%%%%%%%%%%%%%%%%
We note that POD-DEIM formulation is independent of the full model dimension $n$ and that the DEIM procedure exploits the structure of the nonlinear function $\vf_w$ as component-wise operation at $\mU_s^r~\tilde{\vy}_s$~\citep{deim}.
%\end{document}
%%%%%%%%%%%%%%%%%%%%%%%%%%%%%%%%%%%%%%%%%
\section{Deep Residual RNN}
\label{drrnnsec}
%%%%%%%%%%%%%%%%%%%%%%%%%%%%%%%%%%%%%%%%%
POD-DEIM reduced order models, as introduced in the last chapter, could be used to perform parametric UQ tasks. However, the POD-DEIM formulation is nonlinear and
% may faces two issues despite the reduction in the dimension of the state variable from $n$ to $r$. The first issue is related to the computational bottle neck in obtaining the solution of the POD-DEIM reduced model, since it
relies on using Newton method at each time step to solve the resulting system of nonlinear equations.
The computational efficiency of the Newton iteration depends on the method employed to assemble the Jacobian matrix and more importantly on the conditioning of the reduced Jacobian matrix. It also depends on the method used to solve the resulting linear system at each iteration of the Newton step and generally, it takes $\mathcal{O}(r^3)$ operations for each saturation update~\citep{2017nagoor, bertsekas1999nonlinear}.
Moreover, previous studies~\citep{ he2013, he2010} pointed to the loss of stability of POD-Galerkin reduced model in several cases and it was attributed to ill-conditioning and poor spectral properties of the reduced Jacobian matrix.

In this paper, we build on the recently introduced \mbox{DR-RNN}~\citep{2017nagoor} and formulate an accurate POD-DEIM reduced order models. \mbox{DR-RNN} is a deep RNN architecture~\citep{2017nagoor}, constructed by stacking $K$ physics aware network layers. \mbox{DR-RNN} could be applied to any nonlinear dynamical system of the form
%%%%%%%%%%%%%%%%%%%%%%%%%%%%%%%%%%%%%%%%%%%%%%%%
\begin{equation}
\dfrac{d\vy}{dt} = \mA~\vy + \vF(\vy)
\label{ODEFOM}
\end{equation}
%%%%%%%%%%%%%%%%%%%%%%%%%%%%%%%%%%%%%%%%%%%%%%%%%%
where ${\vy}(\va, t)\in \mathbb{R}^n$ is the state variable at time $t$, $\va \in \mathbb{R}^d$ is a system parameter vector, the matrix $\mA \in \mathbb{R}^{n\times n}$ is the linear part of the dynamical system and the vector $\vF({\vy}) \in \mathbb{R}^n$ is the nonlinear term~\citep{2017nagoor}.
The state variable $\vy(t)$ at different time steps is obtained by solving the nonlinear residual equation defined as
%%%%%%%%%%%%%%%%%%%%%%%%%%%%%%%%%%%%%%%%%%%%%%%%%%%%%%%%%%%%%%%%%%%%%%%%%%%%%%%%%%
\begin{equation}
\vr_{t+1} = \vy_{t+1} - \vy_t - \Delta t~\mA~\vy_{t+1} - \Delta t~\vF(\vy_{t+1})
\label{timeresidualEuler}
\end{equation}
%%%%%%%%%%%%%%%%%%%%%%%%%%%%%%%%%%%%%%%%%%%%%%%%%%%%%%%%%%%%%%%%%%%%%%%%%%%%%%%%%%%%%%%%
where $\vr(t)$ is termed as the residual vector at time step $t$ and $\vy(t+1)$ is the approximate solution of Eq.~\eqref{ODEFOM} at time step $t+1$ obtained by using implicit Euler time integration method. \mbox{DR-RNN} ~\citep{2017nagoor} approximates the solution of Eq.~\eqref{ODEFOM} using the following iterative update equations
%%%%%%%%%%%%%%%%%%%%%%%%%%%%%%%%%%%%%%%%%%%%%%%%%%%%%%%%%%%%%%%%%%%%%%%%%%%%%%%%%%%%%%%%%%%%%%%%%%%%%%%%%%
\begin{equation}
\begin{aligned}
\vy^{(k)}_{t+1} &= \vy^{(k-1)}_{t+1} - \vw~\circ~\phi_h(\mU~\vr^{(k)}_{t+1}) &\qquad\text{for}~k = 1, \\
\vy^{(k)}_{t+1} &= \vy^{(k-1)}_{t+1} - \frac{\eta_k}{\sqrt{G_k+\epsilon}}~\vr^{(k)}_{t+1} &\qquad\text{for}~k > 1,
\label{DR-RNNequation}
\end{aligned}
\end{equation}
%%%%%%%%%%%%%%%%%%%%%%%%%%%%%%%%%%%%%%%%%%%%%%%%%%%%%%%%%%%%%%%%%%%%%%%%%%%%%%%%%%%%%%%%%%%%%%%%%%%%%%%%%%%%%
where $\mU, \vw, \eta_k$ are the training parameters of DR-RNN, $\phi_h$ is the $\tanh$ activation function, $\circ$ is an element-wise multiplication operator, $\vr^{(k)}_{t+1}$ is the residual in layer $k$ obtained by substituting $\vy_{t+1} = \vy^{(k-1)}_{t+1}$ into Eq.~\eqref{timeresidualEuler} and $G_k $ is an exponentially decaying squared norm of the residual defined by
%%%%%%%%%%%%%%%%%%%%%%%%%%%%%%%%%%%%%%%%%%%%%%%%%%%%%%%%%%%%%
\begin{equation}
G_{k} = \gamma~\Vert\vr^{(k)}_{t+1}\Vert^2 + \zeta~G_{k-1}
\label{rmspropDR-RNN}
\end{equation}
%%%%%%%%%%%%%%%%%%%%%%%%%%%%%%%%%%%%%%%%%%%%%%%%%%%%%%%%%%%%%%
where $\gamma, \zeta$ are fraction factors and $\epsilon$ is a smoothing term to avoid divisions by zero~\citep{2017nagoor}.
In this formulation, we set $\vy^{(k=0)}_{t+1} = \vy_t$. The architecture of \mbox{DR-RNN} is inspired by the rmsprop algorithm~\citep{tieleman2012lecture} which is a variant of the steepest descent method. The \mbox{DR-RNN} output at each time step is defined as
%%%%%%%%%%%%%%%%%%%%%%%%%%%%%%%%%%%%%%%%%%%%%%%%%%%%%
\begin{equation}
\vy_{t+1}^{\text{\tiny{(RNN)}}} = \vy_{t+1}^K
\end{equation}
%%%%%%%%%%%%%%%%%%%%%%%%%%%%%%%%%%%%%%%%%%%%%%%%%%%%%
The formulation of \mbox{DR-RNN} is explicit in time and has a fixed number of iterations $K$ per time step. However, the dimension of the \mbox{DR-RNN} system depends on the dimension of the residual. For example, \mbox{DR-RNN} (Eq.~\eqref{DR-RNNequation}) can be derived from the POD-DEIM reduced model residual ($\tilde{\vr}_{t+1} = -\tilde{\vy}_{s_{t+1}} + \tilde{\vy}_{s_{t}} + \mD~\vf_w(\mP^{\top}~\mU_{s}^r~\tilde{\vy}_{s_{t+1}}) + \tilde{\vd}$).
In such setting, the \mbox{DR-RNN} dynamics has a fixed computational budget of $\mathcal{O}(r^2)$ for each time step. In addition, \mbox{DR-RNN} has the prospect of employing large time step violating the numerical stability constraint~\citep{2017nagoor}. Furthermore, \mbox{DR-RNN} does not rely on the reduced Jacobian matrix to approximate the solution of POD-DEIM reduced model.

The \mbox{DR-RNN} parameters $\TT = \lbrace \mU,~\vw,~\eta_k \rbrace$
are fitted by minimizing the mean square error (mse) defined by
%%%%%%%%%%%%%%%%%%%%%%%%%%%%%%%%%%%%%%%%%%%%%%%%%%%%%%%%%%%%%%%%%%%%%%%%%%%%%%%%%%%%%%%%%%%%%%%
\begin{align}
\jacob_{\text{\tiny{MSE}}}(\TT) = \frac{1}{L}\sum_{\ell=1}^L \sum_{t=1}^{T} ( \vy_{t} - \vy_{t}^{\text{\tiny{(RNN)}}} )^2 ,
\label{mseloss}
\end{align}
%%%%%%%%%%%%%%%%%%%%%%%%%%%%%%%%%%%%%%%%%%%%%%%%%%%%%%%%%%%%%%%%%%%%%%%%%%%%%%%%%%%%%%%%%%%%%%%%
where $\jacob_{\text{\tiny{MSE}}}$ (mse) is the average distance between the reference solution $\vy_{t}$ and the RNN output ${\vy}_{t}^{\text{\tiny{RNN}}}$ across a number of samples $L$ with time-dependent observations $(t=1~\cdots~T~\text{and}~\ell=1~\cdots~L)$~\citep{2017nagoor, pascanu2013construct}.
The set of parameters $\TT$ is commonly estimated by a technique called Backpropagation Through Time (BPTT)~\citep{werbos1990backpropagation,rumelhart1986learning,pascanu2013difficulty,mikolov2014learning}, which backpropagates the gradient of the loss function $\jacob_{\text{\tiny{MSE}}}$ with respect to $\TT$ in time over the length of the simulation.

%%%%%%%%%%%%%%%%%%%%%%%%%%%%%%%%%%%%%%%%%%%%%%%
\section{Numerical Experiments}
\label{numericalexamples}
%%%%%%%%%%%%%%%%%%%%%%%%%%%%%%%%%%%%%%%%%%%%%%%%%%
In this section, we evaluate the performance of the reduced order models based on \mbox{DR-RNN} against the standard implementation of POD-Galerkin reduced model. Specifically, we develop two POD-Galerkin based reduced model using \mbox{DR-RNN} architecture namely, \mbox{DR-RNN}$^{\text{p}}$ (\mbox{DR-RNN} combined with POD-Galerkin) and \mbox{DR-RNN}$^{\text{pd}}$ (\mbox{DR-RNN} combined with POD-Galerkin and DEIM).
The numerical evaluations are performed using two uncertainty quantification tasks involving subsurface flow models. We did not include %a
~standard POD-DEIM reduced model implementation as we expect that the standard POD reduced model results to be far superior~\citep{deim, 2017nagoor, deim}.

The outline of this section is as follows: In subsection~\ref{FOMsetup}, we present the description of the flow problem, followed by a brief description of the finite-volume approach employed for obtaining the full-order model solution. Following that, in subsection~\ref{PODROMsetup}, we outline the specific details to formulate POD reduced model. Then, we list the settings adopted to model the \mbox{DR-RNN} ROMs (i.e. number of layers, optimization settings, etc) in the subsection~\ref{drrnnsetup}. In subsection~\ref{errormetrics}, we provide a set of error metrics utilized to evaluate the performance of the different ROMs. In subsection~\ref{problem1}, we present the numerical results for the quarter five spot model followed by results for the uniform flow model in the subsection~\ref{problem2}.

%%%%%%%%%%%%%%%%%%%%%%%%%%%%%%%%%
% \subsection{Problem description}
% \label{problemdescription}
%%%%%%%%%%%%%%%%%%%%%%%%%%%%%%%%%%%
\subsection{Full-order model setup}
\label{FOMsetup}
\begin{figure}[h!]
\centering
\includegraphics[width=1.1\textwidth]{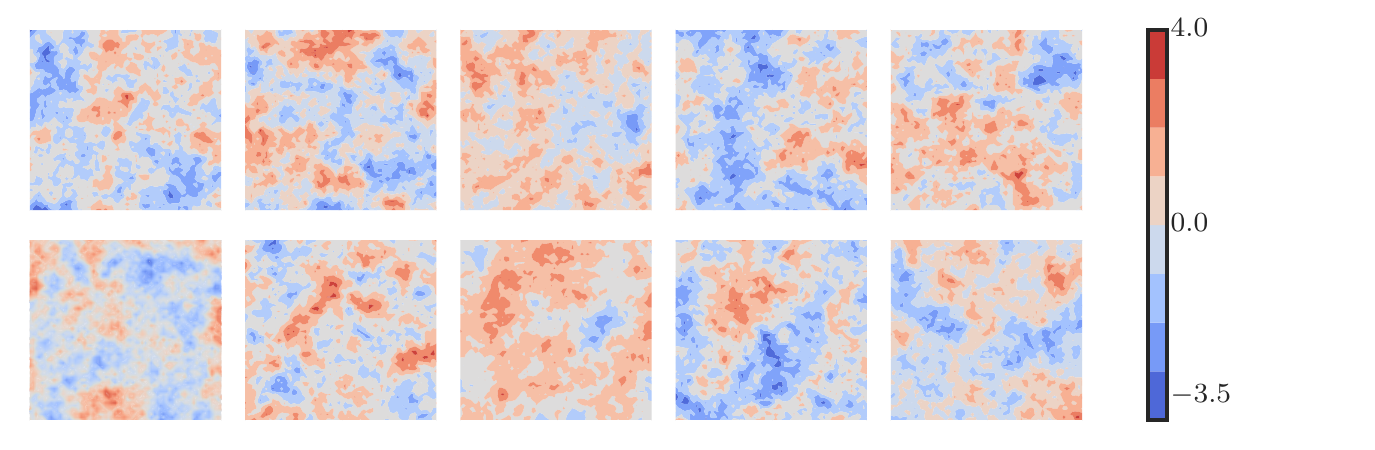}
\caption{Plots of log values of random permeability field modeled by log-normal probability distribution. The unit of the permeability field is $m^2$.}
\label{lognormalfig}
\end{figure}

%%%%%%%%%%%%%%%%%%%%%%%%%%%%%%%%%%%%
We consider a two-phase (oil and water) porous media flow problem over the two-dimensional domain $[0~1] \times [0~1]$ meters. The equations governing the two-phase flow are the pressure equation (Eq.~\eqref{pressure_eq}) and the saturation equation (Eq.~\eqref{saturation_eq}).
The relative permeability is defined as a function of saturation using Corey's model $k_{rw}(s)={s^*}^2$, $k_{ro}=(1-s^*)^2$, where $s^*=(s-s_{wc})/(1-s_{or}-s_{wc})$, $s_{wc}$ is the irreducible water saturation and $s_{or}$ is the residual oil saturation~\citep{aarnes2007introduction}.
We set $s_{or}=0.2$ and $s_{wc}=0.2$. We set the initial water saturation over the domain to the irreducible water saturation $s_{wc}=0.2$.
The water and oil viscosities are 1 and 1.5 centipoise, respectively. The porosity is assumed to be a constant value of 0.2 over the entire problem domain. The uncertain permeability field is modeled as a log-normal distribution function with zero mean and exponential covariance kernel of the form
%%%%%%%%%%%%%%%%%%%%%%%%%%%%
\begin{equation}
\mathbb{C}ov = \sigma_k~\exp \left[ -\frac{\vert x_1 - x_2 \vert }{L_k} \right]
\label{covariance}
\end{equation}
%%%%%%%%%%%%%%%%%%%%%%%%%%%%%%
where $\sigma_k$ is the variance, $L_k$ is the correlation length.
In all test cases, we set $\sigma_k$ to 1 and the correlation length $L_k$ to $0.1$. Figure~\ref{lognormalfig} shows several realizations of the log-permeability values. %%%%%%%%%%%%%%%%%%%%%%%%%%%
%%%%%%%%%%%%%%%%%%%%%%%%%%%%%%%%%
For solving the full-order model, the problem domain is discretized using a uniform grid of $64 \times 64$ blocks. The pressure equation is discretized using simple finite volume method (aka. Two Point Flux Approximation)~\citep{aarnes2007introduction} and an upwind finite-volume scheme is used to discretized the saturation equation. For the time discretization, an implicit backward Euler method combined with Newton-Raphson iterative method is used to solve the resulting system of nonlinear equations. We set the time step size to $0.015$ and the total number of time steps is set to $160$. We note that, the time is measured in a non-dimensional quantity called pore volumes injected (PVI).
PVI defines the net volume of water injected as a fraction of the total pore volume.
As the pressure changes at much slower rate than the saturation, the pressure equation (and hence the velocity) is solved at every 8th saturation time step. For reference solutions, this system of equations is solved for $2000$ random permeability realizations to estimate an ensemble based statistics using Monte-Carlo method~\citep{ibrahima2016}.

\subsection{POD-Galerkin based reduced model setup}
\label{PODROMsetup}
The first step in formulating POD reduced model is to compute the optimal POD basis matrices $\mU_p^r$ and $\mU_s^r$. In order to obtain these basis matrices, we initially preformed a realization clustering algorithm to enforce the diversity of the collected snapshots and clustered the $2000$ random permeability realizations into $45$ clusters~\citep{gasemi2015}. Then, we randomly selected a single permeability realization from each cluster (total 45 random samples of the permeability field). The full system is then solved for each of the 45 realizations and the solution vectors are collected to build the snapshot matrices (pressure, saturation, nonlinear function). Finally, we compute the POD basis matrices from the SVD of the collected snapshot matrices.

Following that, the obtained basis vectors are used to build POD reduced model (as detailed in the section~\ref{reducedordermodel}). We then employ the same sequential implicit technique settings adopted for obtaining the full model solutions to solve the resultant POD reduced model. For numerical evaluations, we solve the POD reduced model for the same $2000$ permeability realizations to estimate an ensemble based statistics in the engineering quantities of interest.

\subsection{\mbox{DR-RNN} setup}
\label{drrnnsetup}
%We now formulate the \mbox{DR-RNN} with six network layers ($K=6$ in Eq.~\eqref{DR-RNNequation}).
In all the numerical test cases, we utilize \mbox{DR-RNN} with six layers ($K=6$ in Eq.~\eqref{DR-RNNequation}).
% to approximate POD-DEIM reduced model (Eq.~\eqref{ROMDEIM_eq}).
We evaluate \mbox{DR-RNN}$^{\text{p}}$ and \mbox{DR-RNN}$^{\text{pd}}$ for different number of POD basis, however, we fix the number of DEIM basis to $35$. The \textsf{PyTorch} framework~\citep{paszke2017automatic}, a deep learning python package using~\textsf{Torch} library as a backend, is used to implement the \mbox{DR-RNN}. Further, we optimize the \mbox{DR-RNN} model parameters using rmsprop algorithm~\citep{tieleman2012lecture,paszke2017automatic} as implemented in \textsf{PyTorch}, where we set the weighted average parameter to $0.9$ and the learning rate to $0.001$. The weight matrix $\mU$ in Eq.~\eqref{DR-RNNequation} is initialized randomly from the uniform distribution function $\mathtt{U [0.01, 0.02]}$. The vector training parameter $\vw$ in Eq.~\eqref{DR-RNNequation} is initialized randomly from the uniform distribution function $\mathtt{U [0.1, 0.5]}$. The scalar training parameters $\eta_k$ in Eq.~\eqref{DR-RNNequation} are initialized randomly from the uniform distribution $\mathtt{U [0.1, 0.4]}$. We set the hyperparameters $\zeta$ and $\gamma$ in Eq.~\eqref{rmspropDR-RNN} to $0.9$ and $0.1$, respectively.
The formulated \mbox{DR-RNN}$^{\text{p}}$ and \mbox{DR-RNN}$^{\text{pd}}$ are trained to approximate the reduced state vector representation obtained from least--squares fits. Specifically, we collect a set of best reduced state vector representation $\tilde{\vy}_s^*$ of the saturation state vector using $\tilde{\vy}_s^* = {\mU_s^r}^{\top}~\vy_s$. The collected set of reduced state vectors is then used to train the parameters of the \mbox{DR-RNN} by minimizing the loss function defined in Eq.~\eqref{mseloss}.

\subsection{Evaluation metrics}
\label{errormetrics}
We evaluate the performance of \mbox{DR-RNN}$^{\text{p}}$ and \mbox{DR-RNN}$^{\text{pd}}$ using two time specific error metrics defined by
%\fbox{\parbox{0.9\linewidth}{%
\begin{equation}
\begin{aligned}
L_{2_{l,t}} &= \Vert \left( \vy_{t} - \vy_{t}^{\text{\tiny{(RM)}}} \right)^l \Vert_2 \\
L_{\infty_{l,t}} &= \Vert \left( \vy_{t} - \vy_{t}^{\text{\tiny{(RM)}}} \right)^l \Vert_{\infty}
\end{aligned}
\label{errorestimators}
\end{equation}%}}
where $l$ is the realization index, and $\vy_{t}^{\text{\tiny{(RM)}}}$ is computed from the reduced model. Additionally, we utilize two relative error metrics defined as
\begin{equation}
\begin{aligned}
L_2^{\text{{\tiny rel}}} &= \frac{1}{L \times T}\sum_{\ell=1}^L \sum_{t=1}^{T} \left \Vert \left( \frac{ \vy_{t} - \vy_{t}^{\text{\tiny{(RM)}}} } {\vy_{t}} \right)^l \right \Vert_2 \\
L_{2\text{,{\tiny max}}}^{\text{{\tiny rel}}} &= \max \limits_{l,t=1~\text{to}~L,T} \left \Vert \left( \frac{ \vy_{t} - \vy_{t}^{\text{\tiny{(RM)}}} } {\vy_{t}} \right)^l \right \Vert_2
\end{aligned}
\label{errorestimators_rel}
\end{equation}%}}
where all the time snapshots of saturation vectors in all realizations are used.

\subsection{Numerical test case 1}
\label{problem1}
\begin{figure}[h!]
\centering
\includegraphics[width=0.95\textwidth]{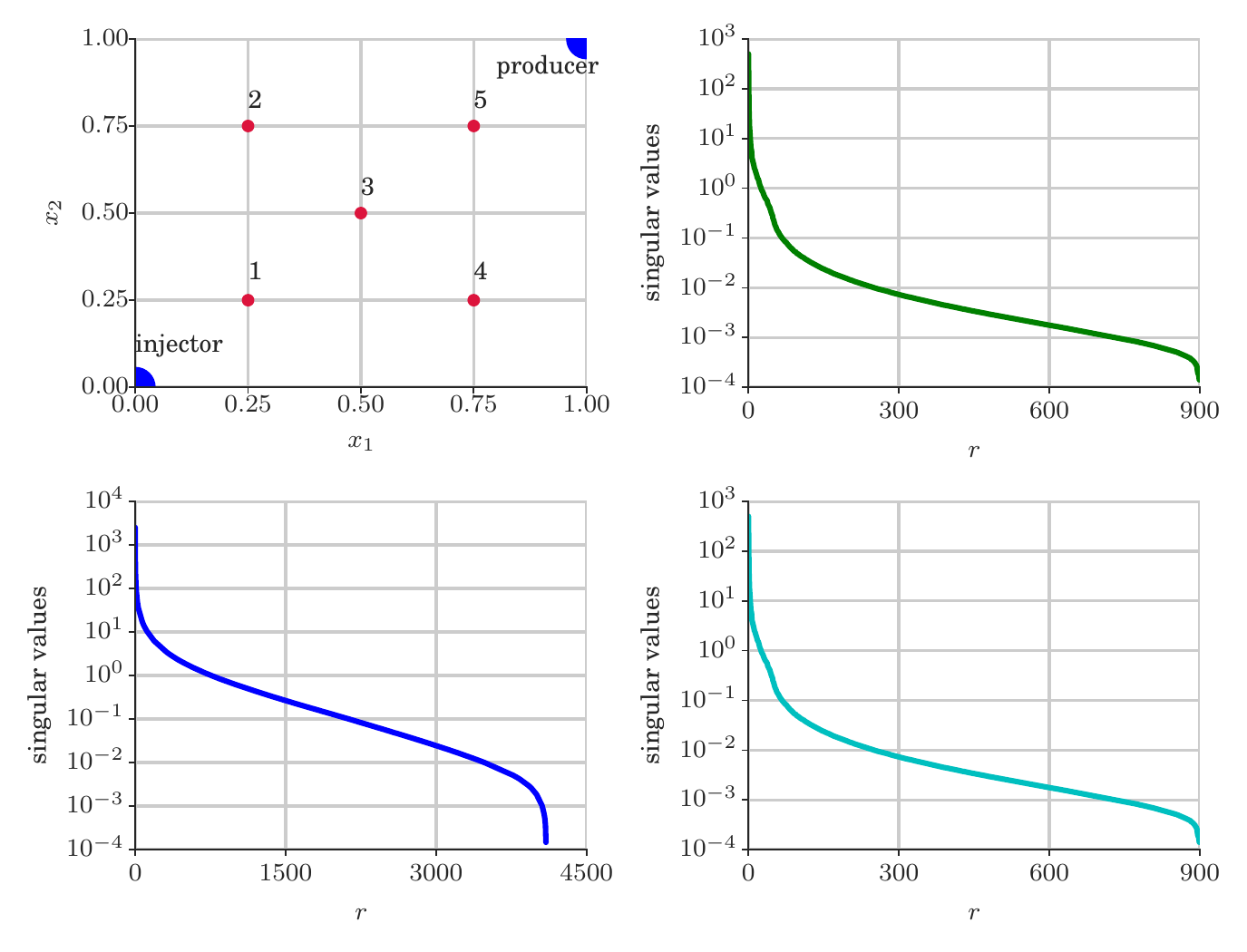}
\caption{Top Left: Computational porous media domain in test case 1. The blue dot in the lower left corresponds to the injector well and the blue dot in the upper right corner corresponds to the
production well. The red dots represented in numbers from 1 to 5 corresponds to the locations where the PDF and the water saturation are investigated. Top Right: Singular values of the pressure snapshot matrix $\mX_p$. Bottom Left: Singular values of the saturation snapshot matrix $\mX_s$. Bottom Right: Singular values of the nonlinear function snapshot matrix $\mX_f$}
\label{singularvalues_p1}
\end{figure}

In this test case, water is injected at the lower left corner ($0,0$) of the domain and a mixture of oil and water is produced at the top right corner of the domain ($1,1$). We set the injection rate $q=0.05$ at ($0,0$) and $q=-0.05$ at ($1,1$) as defined in Eq.~\eqref{saturation_eq}. We impose a no flow boundary condition in all the four sides of the domain. We fix the number of pressure POD basis to 5 and obtain all the ROMs for a set of different number of saturation POD basis functions ($r=10, 20$).
The configuration of the problem domain is shown in top left panel of Figure~\ref{singularvalues_p1}, where the blue spot in the lower left corner (0,0) corresponds to the injector well and the blue spot in the upper right corner (1,1) corresponds to the production well. Figure~\ref{singularvalues_p1} shows the singular values of the pressure snapshot matrix $\mX_p$ in the top right panel, the saturation snapshot matrix $\mX_s$ in the bottom left panel, and the nonlinear function snapshot matrix $\mX_f$ in the bottom right panel.
%%%%%%%%%%%%%%%%%%%%%%%%%%%
\begin{figure}[h!]
\centering
\includegraphics[width=0.95\textwidth]{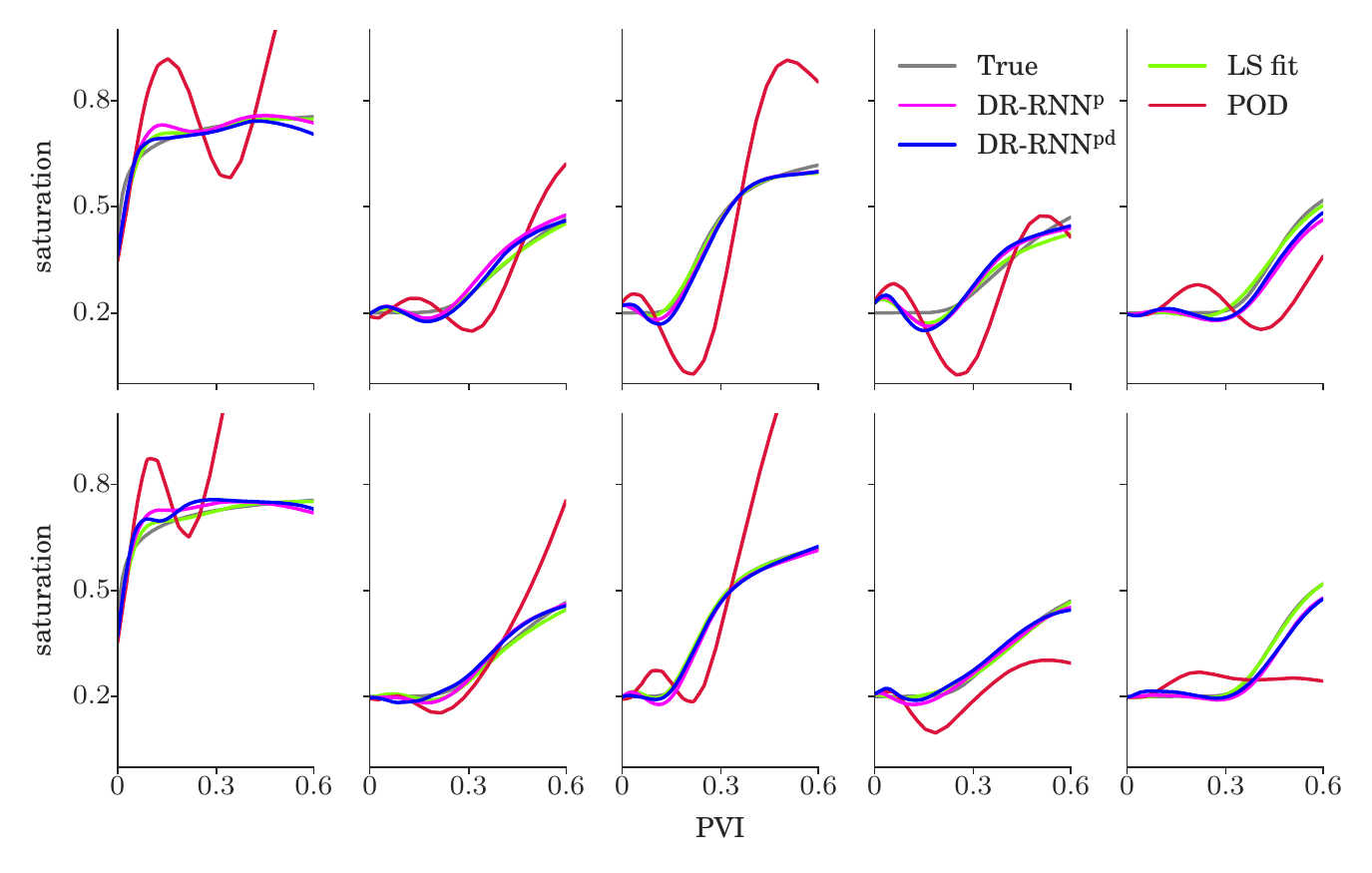}
\caption{Time plots of mean water saturation obtained from all the ROMs and the full-order model for test case 1. Top Row: number of POD basis used $=10$. Bottom Row: number of POD basis used $=20$. The plots in each row are arranged as per the numerical notation of the spatial points plotted in Figure~\ref{singularvalues_p1} (top left panel).}
\label{watercutfig}
\end{figure}
%%%%%%%%%%%%%%%%%%%%%%%%%%%

The mean water saturation plots over the simulation time are shown in Figure~\ref{watercutfig}, where the results in the top row corresponds to using $10$ POD basis and the results in the bottom row corresponds to using $20$ POD basis. The sub-plots in Figure~\ref{watercutfig} are arranged from left to right following the numbering of the spatial points shown in Figure~\ref{singularvalues_p1}. From these results, it is clear that DR-DR-RNN$^{\text{p}}$ and \mbox{DR-RNN}$^{\text{pd}}$ results are very close to the least--square solutions (LS fit). In Figure~\ref{watercutfig}, POD-Galerkin reduced model yields extremely inaccurate and unstable results. We attribute the small errors in \mbox{DR-RNN}$^{\text{p}}$ and \mbox{DR-RNN}$^{\text{pd}}$ results to the insufficient number of POD basis vectors and we note that the error magnitude is equivalent to the optimal values obtained by least-squares projection.
%%%%%%%%%%%%%%%%%%%%%%%%%%%
%\end{document}
\begin{figure}[h!]
\centering
\includegraphics[width=1.1\textwidth]{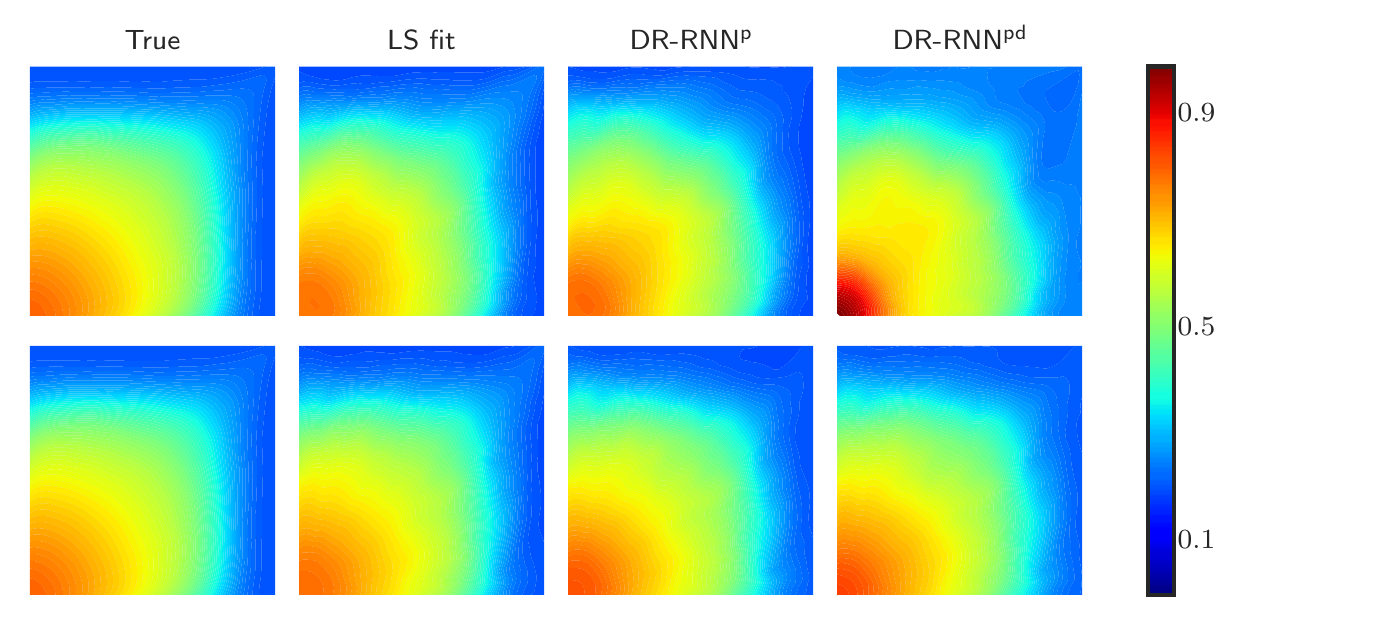}
\caption{Comparison of mean water saturation field at time $ = 0.3$ PVI for test case 1. Top Row: number of POD basis used $=10$. Bottom Row: number of POD basis used $=20$.}
\label{saturationmeanfig}
\end{figure}
%\end{document}
\begin{figure}[h!]
\centering
\includegraphics[width=1.1\textwidth]{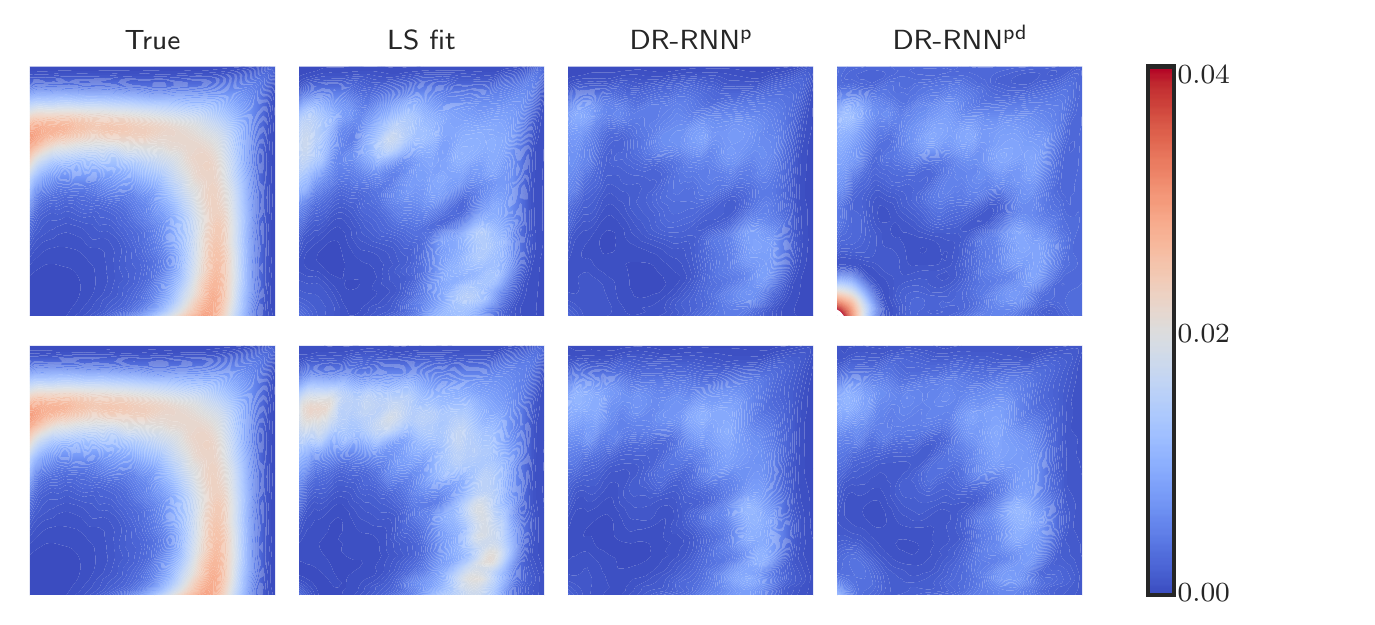}
\caption{Comparison of standard deviation of the water saturation field at time $= 0.3$ PVI for test case 1. Top Row: number of POD basis used $=10$. Bottom Row: number of POD basis used $=20$.}
\label{msefig}
\end{figure}
%%%%%%%%%%%%%%%%%%%%%%%%%%%

\begin{figure}[h!]
\centering
\includegraphics[width=0.8\textwidth]{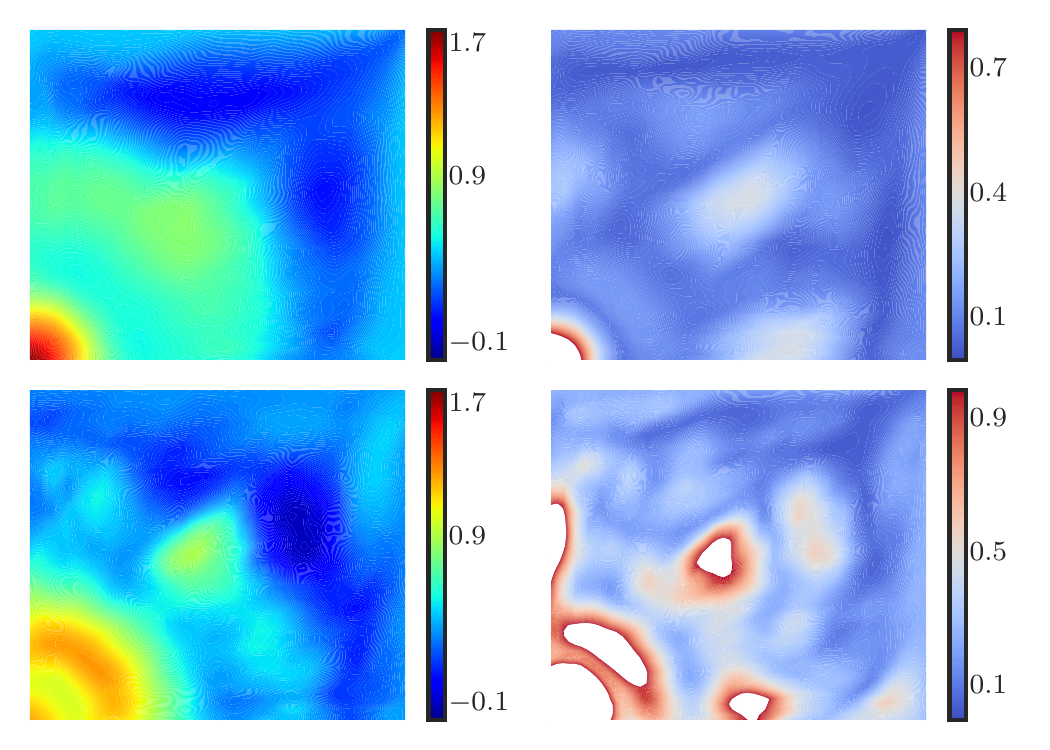}
\caption{Plot of saturation mean and standard deviation of the water saturation field at time $= 0.3$ PVI obtained from the POD reduced model for test case 1. Left: saturation mean. Right: standard deviation. Top Row: number of POD basis used $=10$. Bottom Row: number of POD basis used $=20$.}
\label{podmsefig}
\end{figure}

Figures~\ref{saturationmeanfig},~\ref{msefig}, and~\ref{podmsefig} show the results for the first (mean) and second (standard deviation) moments of the saturation field at time $= 0.3$ PVI obtained from the full model and from the various ROMs. In these figures (\ref{saturationmeanfig},~\ref{msefig}, and~\ref{podmsefig}), results for 10 POD basis are shown in the top row and results for 20 POD basis are shown in the bottom row. As shown in Figure~\ref{saturationmeanfig}, the mean saturation obtained from \mbox{DR-RNN} ROMs are almost indistinguishable from the reference results. However, the mean saturation field obtained from POD reduced model (left panels of Figure~\ref{podmsefig}) deviates significantly from the reference mean saturation.

In Figure~\ref{msefig}, we observe small discrepancy of standard deviation results obtained the \mbox{DR-RNN} ROMs in comparison to the full model results especially near the location of the mean water saturation front. Figure~\ref{podmsefig} (right panels) shows the standard deviation results obtained by POD reduced model which show significant inaccuracies that could be indicative to instabilities of the obtained solutions. We note that the white spots in Figure~\ref{podmsefig} correspond to out of limits shown in colorbar.
%%%%%%%%%%%%%%%%%%%%%%%%%%%
%\begin{figure}[H]
%\begin{center}
%\includegraphics{ONEl2norm.pdf}
%%\caption{Left mean square error for the pressure state vector. Right mean square error for the saturation state vector. $r$ denotes number of basis vectors.}
%\end{center}
%\label{msefig}
%\end{figure}
%%%%%%%%%%%%%%%%%%%%%%%%%%%
%%%%%%%%%%%%%%%%%%%%%%%%%%%
\begin{figure}[h!]
\centering
\includegraphics[width=0.95\textwidth]{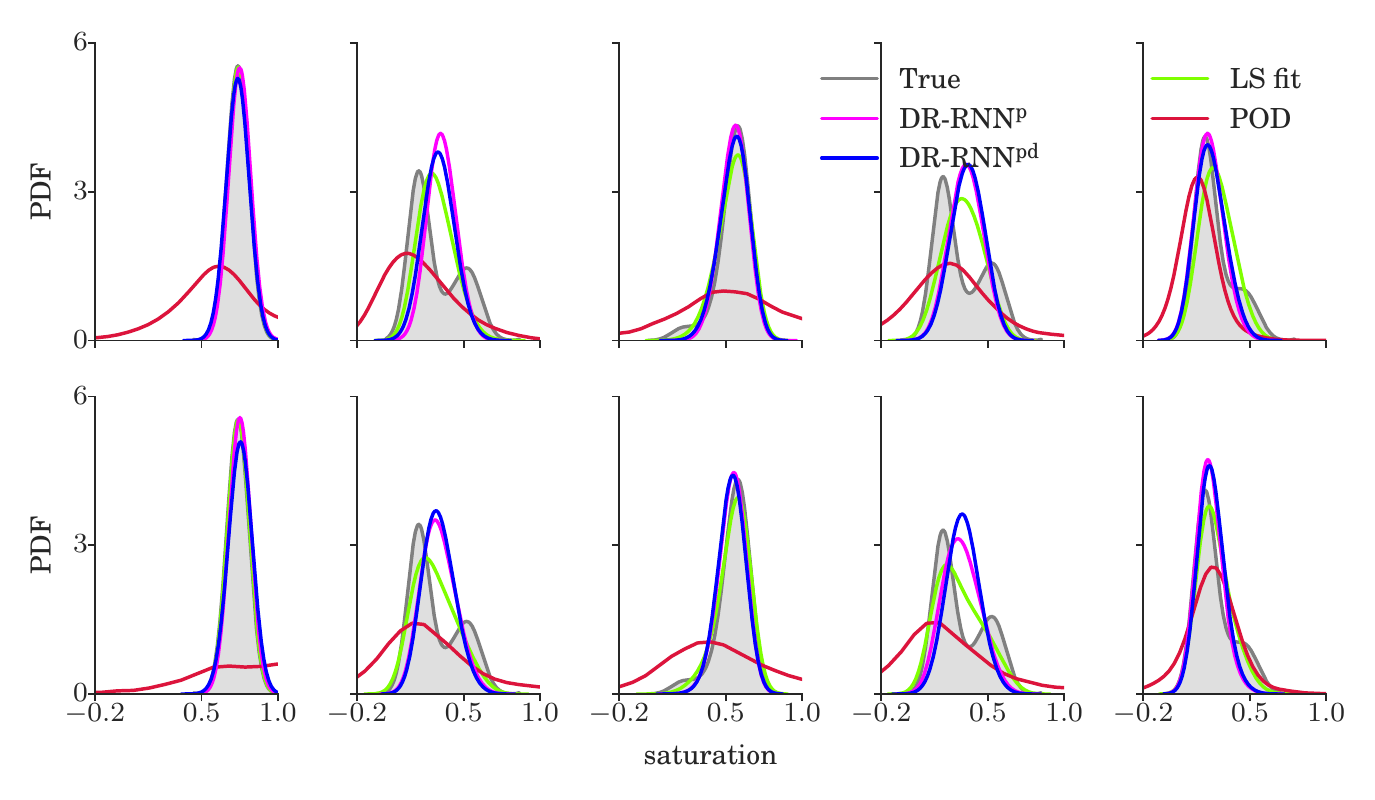}
\caption{Comparison of kernel density estimated probability density function (PDF) at time $= 0.3$ PVI for test case 1. Top Row: number of POD basis used $=10$. Bottom Row: number of POD basis used $=20$. The plots in each row are arranged as per the numerical notation of the spatial points plotted in Figure~\ref{singularvalues_p1} (top left panel).}
\label{kdefig}
\end{figure}
%\begin{figure}[H]
%\begin{center}
%\includegraphics{ONEkdecdfplot.pdf}
%%\caption{Left mean square error for the pressure state vector. Right mean square error for the saturation state vector. $r$ denotes number of basis vectors.}
%\end{center}
%\label{msefig}
%\end{figure}
%%%%%%%%%%%%%%%%%%%%%%%%%%%

Figure~\ref{kdefig} compares the saturation PDF estimated from the ensemble of numerical solutions (ROMs and the full model). Figure~\ref{kdefig} settings are similar to the one adopted in Figure~\ref{watercutfig}. In Figure~\ref{kdefig}, we can see that all the plots obtained from DR-DR-RNN$^{\text{p}}$ and \mbox{DR-RNN}$^{\text{pd}}$ are indistinguishable from the plots obtained from the LS fit (the best approximation). Further, we observe that the saturation PDF obtained from DR-DR-RNN$^{\text{p}}$ and \mbox{DR-RNN}$^{\text{pd}}$ follow nearly the same trend of saturation PDF obtained from the full model when the reference distribution is unimodal. However, we observe some discrepancy when the distributions are multimodal. Please note that similar discrepancy is also observed in the PDF obtained from LS fit. Hence, we postulate that these discrepancies are attributed to the limited number of POD basis vectors utilized. In Figure~\ref{kdefig}, POD reduced model yields very inaccurate approximation of the saturation PDF irrespective of the number of POD basis.
%%%%%%%%%%%%%%%%%%%%%%%%%%%
\begin{figure}[h!]
\centering
\includegraphics[width=0.95\textwidth]{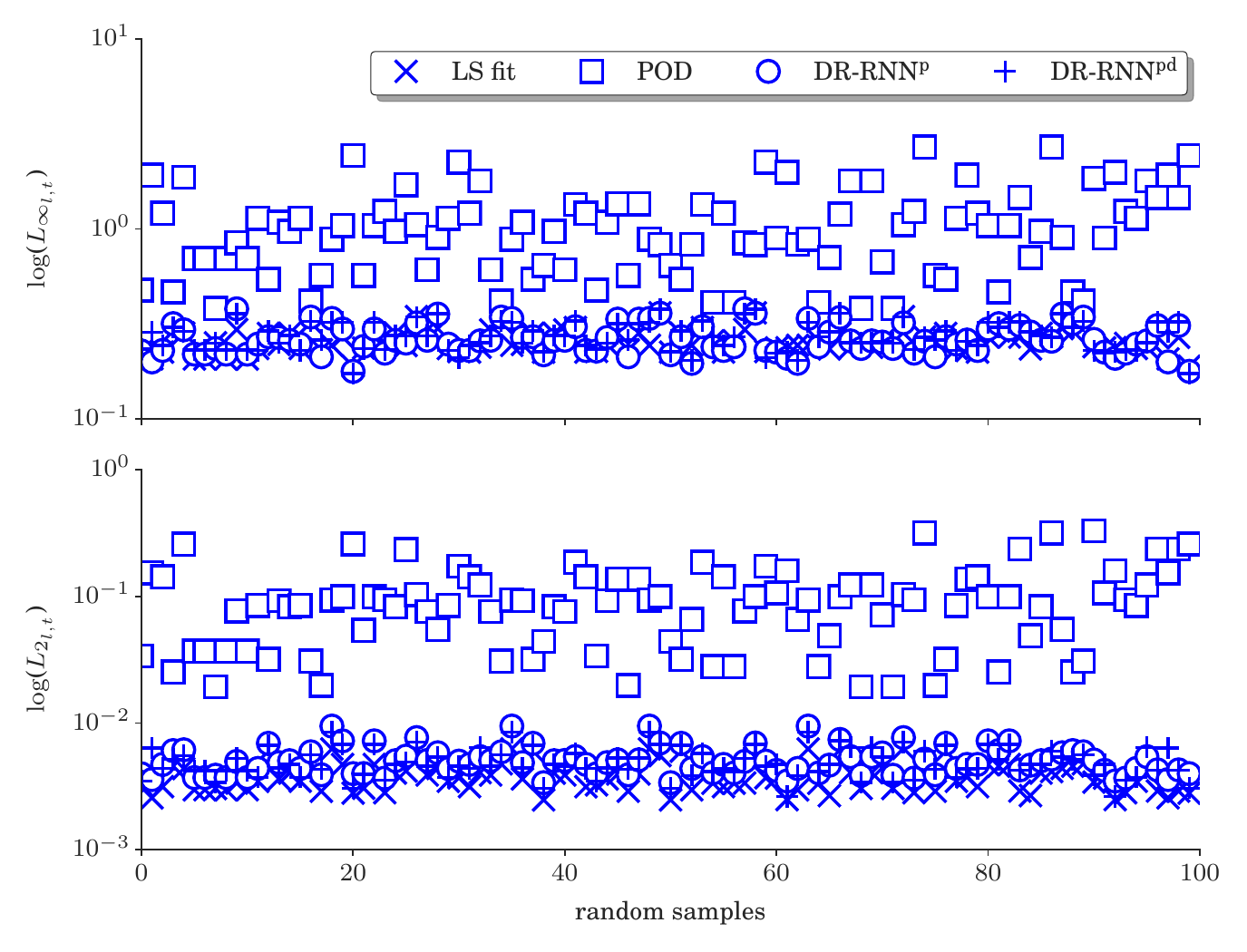}
\caption{Comparison of $\log(L_{2_{l,t}})$ and $\log(L_{\infty_{l,t}})$ error estimators (Eq.~\eqref{errorestimators}) at time $= 0.3$ PVI for test case 1. The number of POD basis used $=10$.}
\label{lnorm10fig}
\end{figure}
\begin{figure}[h!]
\centering
\includegraphics[width=0.95\textwidth]{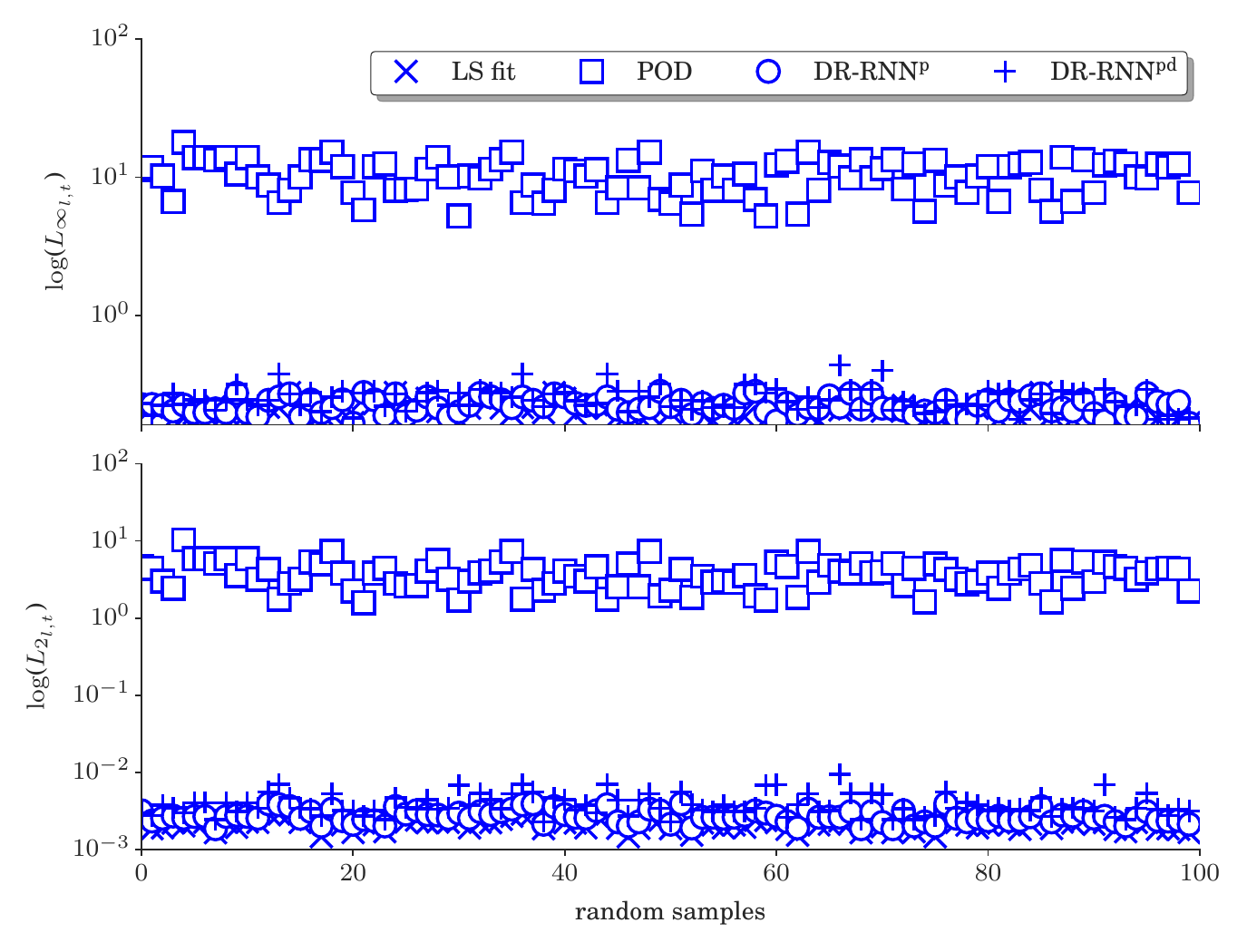}
\caption{Comparison of $\log(L_{2_{l,t}})$ and $\log(L_{\infty_{l,t}})$ error estimators (Eq.~\eqref{errorestimators}) at time $= 0.3$ PVI for test case 1. The number of POD basis used $=20$.}
\label{lnorm20fig}
\end{figure}
%%%%%%%%%%%%%%%%%%%%%%%%%%%
%%%%%%%%%%%%%%%%%%%%%%%%%%%%

Figures~\ref{lnorm10fig} and~\ref{lnorm20fig} displays samples of $\log(L_{2_{l,t}})$ and $\log(L_{\infty_{l,t}})$ errors at time $ 0.3$ PVI obtained from all the ROMs. All the ROMs use 10 POD basis to display the errors in Figure~\ref{lnorm10fig} and likewise 20 POD basis to display the errors in Figure~\ref{lnorm20fig}. From these figures, we can see that the POD reduced model approximation errors are at least an order of magnitude more than the least--squares solution errors (Eq.~\eqref{Least--squaresolution_eq}), whereas the errors obtained from \mbox{DR-RNN}$^{\text{p}}$ and \mbox{DR-RNN}$^{\text{pd}}$ are nearly indistinguishable from the least--squares projection errors.
%%%%%%%%%%%%%%%%%%%%%%%%%%%%%%%%%%%%%%%%%%%%%%%%%%%%%%%%%%%%%%%%%%%%%
\begin{table}[h!]
\caption{Performance chart of all the ROMs employed for test case 1. $L_2^{\text{{\tiny rel}}}$ and $L_{2\text{,{\tiny max}}}^{\text{{\tiny rel}}}$ error estimators are defined in the Eq.~\eqref{errorestimators_rel}. The number of POD basis used $=10$ and $20$.}
\centering
% centering table
\begin{tabular}{l c cccc}
\noalign{\smallskip} \hline
%\hline
\noalign{\smallskip
}
Error & \#Basis &\multicolumn{4}{c}{Reduced Order Models} \\
\noalign{\smallskip} \hline %\hline
\noalign{\smallskip
}
% inserts single-line
% Entering 1 st row
 & &LS fit & POD & \mbox{DR-RNN}$^{\text{p}}$ & \mbox{DR-RNN}$^{\text{pd}}$ \\[0.5ex]
 \hline\noalign{\smallskip
}
% Entering 2 nd row
 &10 & 0.13 & 0.56 & 0.14 & 0.15 \\[-1ex]
 \raisebox{1.5ex}{$L_2^{\text{{\tiny rel}}}$} & 20
 & 0.10 & 2.7 & 0.11 & 0.13 \\[1ex]
% Entering 3 rd row
 &10 & 0.20 & 1.8 & 0.20 & 0.27 \\[-1ex]
 \raisebox{1.5ex}{$L_{2\text{,{\tiny max}}}^{\text{{\tiny rel}}} $} & 20
 & 0.17 & 5.8 & 0.19 & 0.26 \\[1ex]
\hline
% inserts single-line
\end{tabular}
\label{taberror}
\end{table}
%%%%%%%%%%%%%%%%%%%%%%%%%%%%%%%%%%%%%%%%%%%%%%%%%%%%%%%%%%%%%%%%%%%%

We further list in Table~\ref{taberror}, the $L_2^{\text{{\tiny rel}}}$ and $L_{2\text{,{\tiny max}}}^{\text{{\tiny rel}}}$ errors for the saturation field. From Table~\ref{taberror}, we can see that the approximation errors obtained from \mbox{DR-RNN}$^{\text{p}}$ and \mbox{DR-RNN}$^{\text{pd}}$ have the same order of magnitude as the least--squares (best approximation) errors. Further, in Table~\ref{taberror}, the approximation errors obtained from all ROMs except POD reduced model decreases when we increase the number of POD basis. These results conform with the decay of singular values of the saturation snapshot matrix. In Table~\ref{taberror}, the approximation errors obtained from POD reduced model are at least an order of magnitude larger than other methods. Also, we observe that POD reduced model results might be worst when we include more basis functions. These results conform with the results presented in~\citep{he2010}, where it was shown that selecting large number of basis vectors based on singular values may not lead to stable POD-Galerkin reduced model. Further, it was presented in~\citep{he2010} that the relation between the stability property of POD-Galerkin reduced model and the number of basis vectors used in POD-Galerkin projection is somewhat random and that the use of more POD basis vectors do not necessarily lead to improved stability. % (see chapter 4 in~\citep{he2010} for more details).

\subsection{Numerical test case 2}
\label{problem2}
\begin{figure}[h!]
\centering
\includegraphics[width=0.95\textwidth]{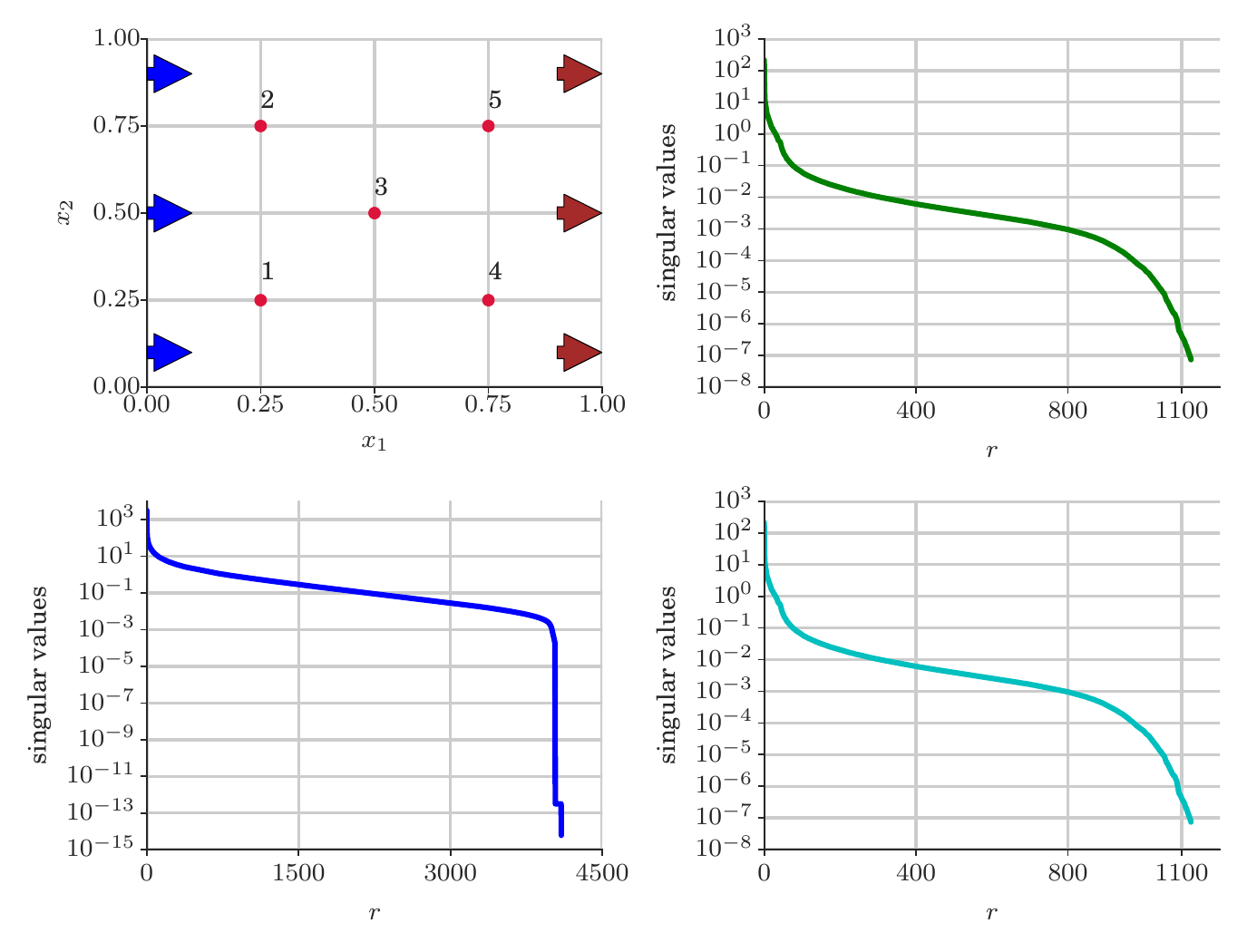}
\caption{Top Left: Computational porous media domain in test case 2. The blue arrows in the left side corresponds to the injection of water and the brown arrows in the right side corresponds to the
production of oil and water. The red dots represented in numbers from 1 to 5 corresponds to the locations where the PDF and the water saturation are investigated. Top Right: Singular values of the pressure snapshot matrix $\mX_p$. Bottom Left: Singular values of the saturation snapshot matrix $\mX_s$. Bottom Right: Singular values of the nonlinear function snapshot matrix $\mX_f$}
\label{singularvalues_p2}
\end{figure}

In this test case, the boundary conditions are set to no flow boundary conditions on the two sides aligned in the horizontal direction (top and bottom). Water is injected from the left side of the domain boundary and fluids are produced from the right side boundary of the domain. The total inflow rate from the left side is set to 0.05 and the total outflow rate from the right side to 0.05 as the problem is incompressible. Similar to test case 1, we evaluate all the ROMs for two different number of saturation POD basis functions ($r=10, 20$). Also, we fix the number of POD basis for the pressure state vector to 5.
Figure~\ref{singularvalues_p2} shows the setup for test case 2 and the corresponding singular values of the snapshot matrices $\mX_p$, $\mX_s$, and $\mX_f$.
%%%%%%%%%%%%%%%%%%%%%%%%%%%
\begin{figure}[h!]
\centering
\includegraphics[width=0.95\textwidth]{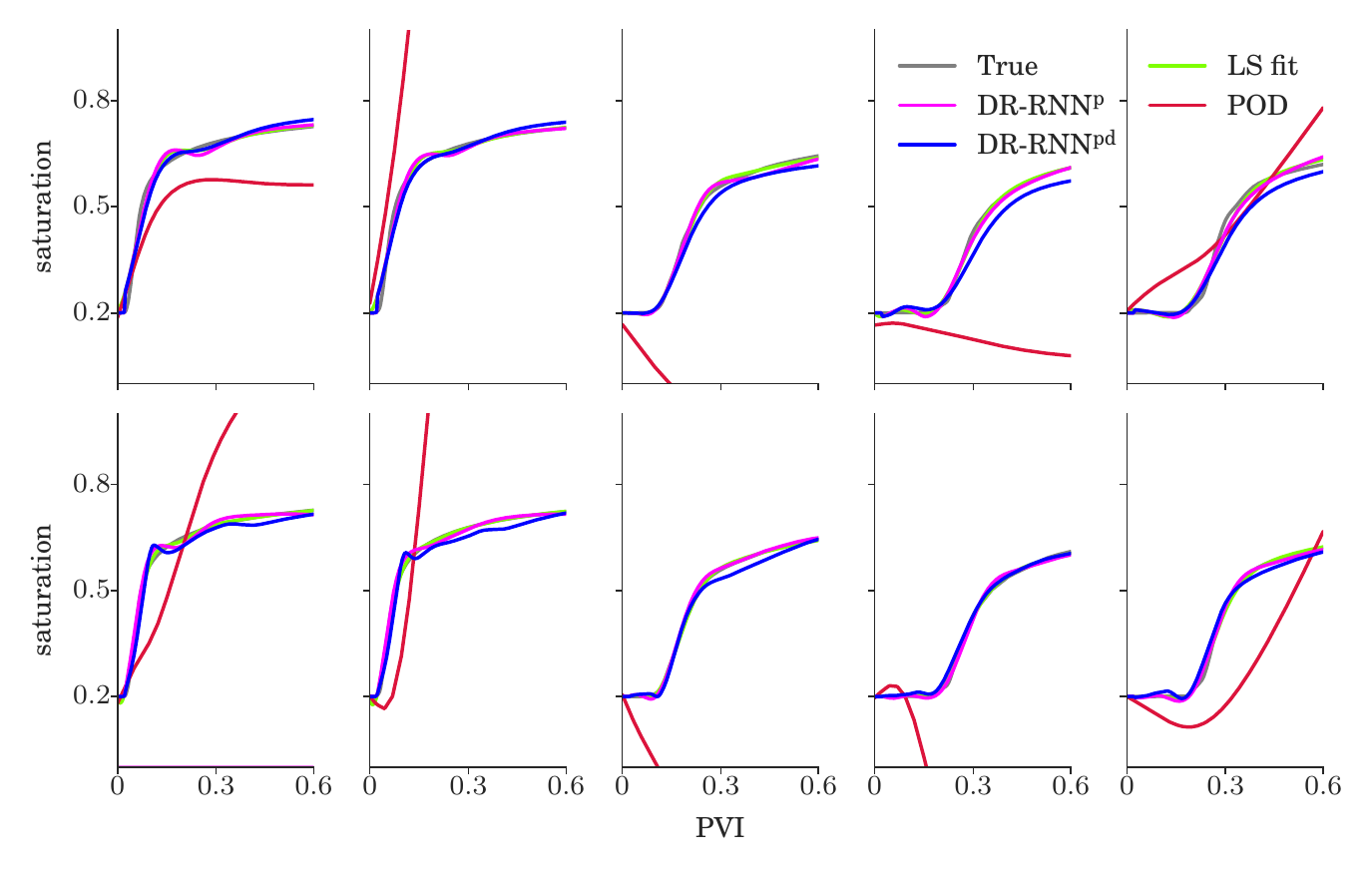}
\caption{Time plots of mean water saturation obtained from all the ROMs and the full-order model in test case 2. Top Row: number of POD basis used $=10$. Bottom Row: number of POD basis used $=20$. The plots in each row are arranged as per the numerical notation of the spatial points plotted in Figure~\ref{singularvalues_p2}.}
\label{uniformwatercutfig}
\end{figure}
%%%%%%%%%%%%%%%%%%%%%%%%%%%

Figure~\ref{uniformwatercutfig} shows the time plot of mean water saturation obtained from all the ROMs and from the full model. The display settings in Figure~\ref{uniformwatercutfig} are the same as defined in Figure~\ref{watercutfig}. In Figure~\ref{uniformwatercutfig}, we can see that all the results obtained from \mbox{DR-RNN}$^{\text{p}}$, \mbox{DR-RNN}$^{\text{pd}}$, and the LS fit (the best approximation) closely approximates the full model whereas POD reduced model yields extremely inaccurate results regardless of the number of utilized POD basis.
%%%%%%%%%%%%%%%%%%%%%%%%%%%
%\end{document}
\begin{figure}[h!]
\centering
\includegraphics[width=1.1\textwidth]{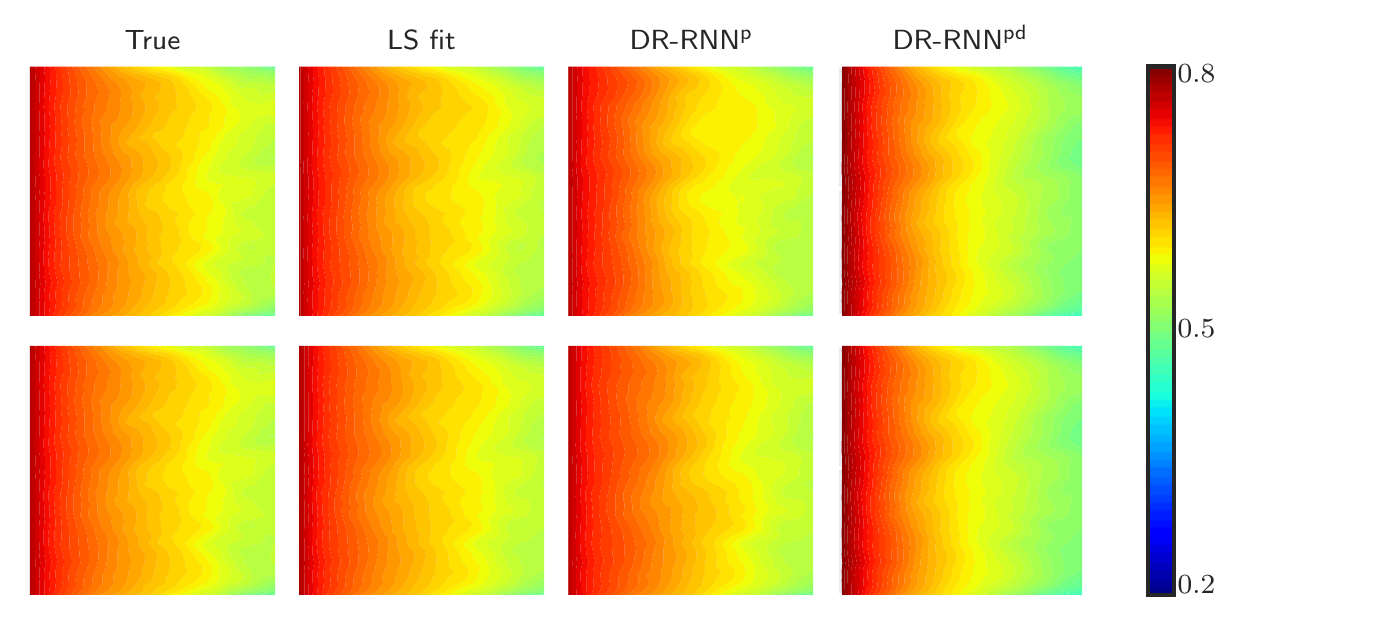}
\caption{Comparison of mean water saturation field at time $ = 0.4$ PVI for test case 2. Top Row: number of POD basis used $=10$. Bottom Row: number of POD basis used $=20$.}
\label{uniformsaturationmeanfig}
\end{figure}
%\end{document}
\begin{figure}[h!]
\centering
\includegraphics[width=1.1\textwidth]{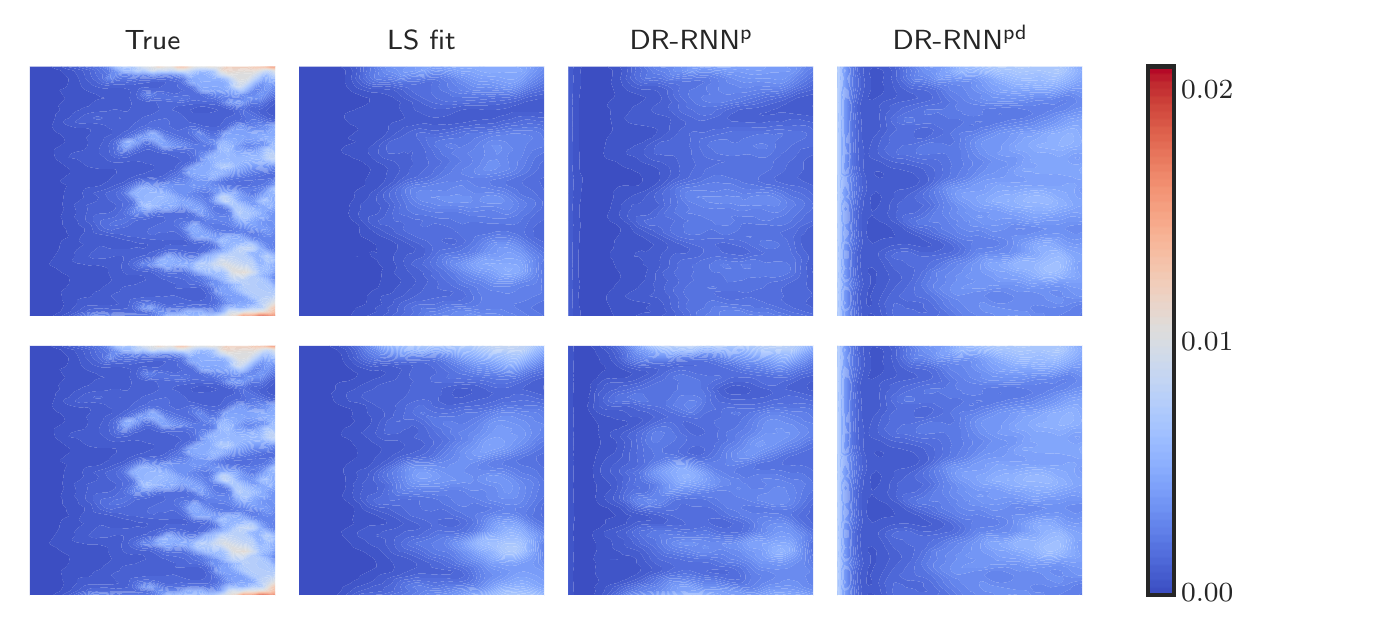}
\caption{Comparison of standard deviation of the water saturation field at time $= 0.4$ PVI for test case 2. Top Row: number of POD basis used $=10$. Bottom Row: number of POD basis used $=20$.}
\label{uniformmsefig}
\end{figure}
%%%%%%%%%%%%%%%%%%%%%%%%%%%

\begin{figure}[h!]
\centering
\includegraphics[width=0.8\textwidth]{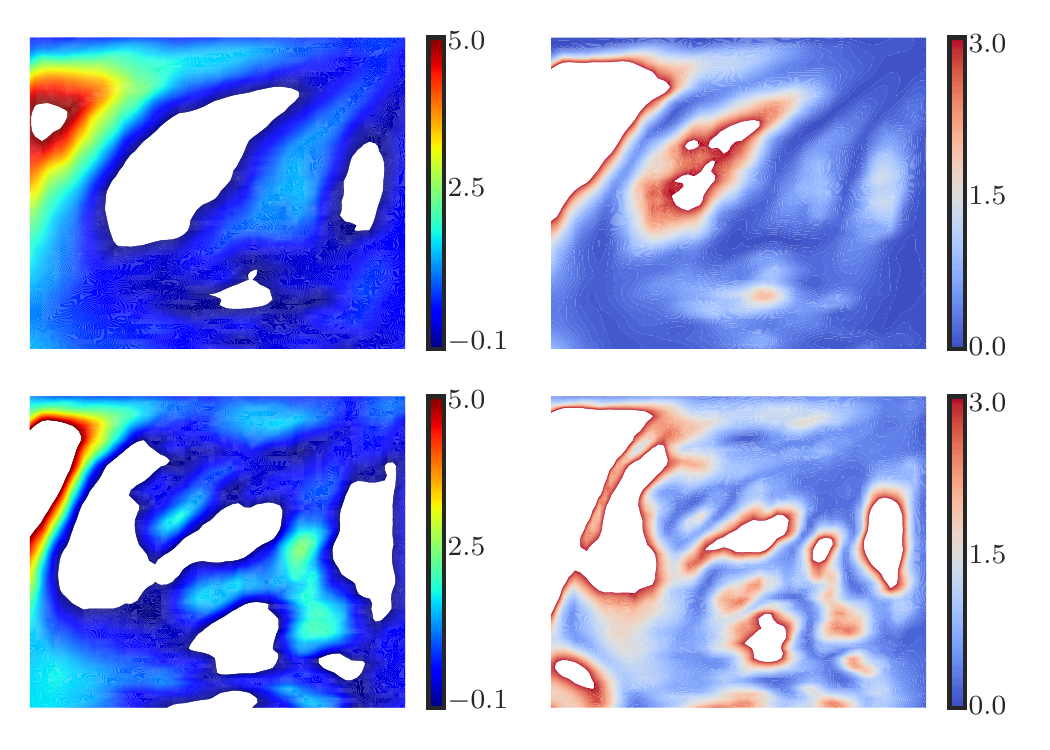}
\caption{Plot of saturation mean and standard deviation of the water saturation field at time $= 0.4$ PVI obtained from the POD reduced model for test case 2. Left: saturation mean. Right: standard deviation. Top Row: number of POD basis used $=10$. Bottom Row: number of POD basis used $=20$.}
\label{uniformpodmsefig}
\end{figure}

Figures~\ref{uniformsaturationmeanfig},~\ref{uniformmsefig}, and~\ref{uniformpodmsefig} shows the results for the mean and standard deviation of the saturation field at $0.4$ PVI.
From these figures, we can conclude that all the plots obtained from \mbox{DR-RNN} ROMs are almost indistinguishable from the LS fit (the best approximation) results, whereas the plots obtained from POD reduced model (Figure~\ref{uniformpodmsefig}) exhibit significant discrepancy when compared to the plots shown in Figure~\ref{uniformsaturationmeanfig}. Again, we note that the white spots displayed in Figure~\ref{uniformpodmsefig} are the regions whose values are out of the limits marked in the respective colorbar.
%%%%%%%%%%%%%%%%%%%%%%%%%%%
%\begin{figure}[H]
%\begin{center}
%\includegraphics{ONEl2norm.pdf}
%%\caption{Left mean square error for the pressure state vector. Right mean square error for the saturation state vector. $r$ denotes number of basis vectors.}
%\end{center}
%\label{msefig}
%\end{figure}
%%%%%%%%%%%%%%%%%%%%%%%%%%%
%%%%%%%%%%%%%%%%%%%%%%%%%%%
\begin{figure}[h!]
\centering
\includegraphics[width=0.95\textwidth]{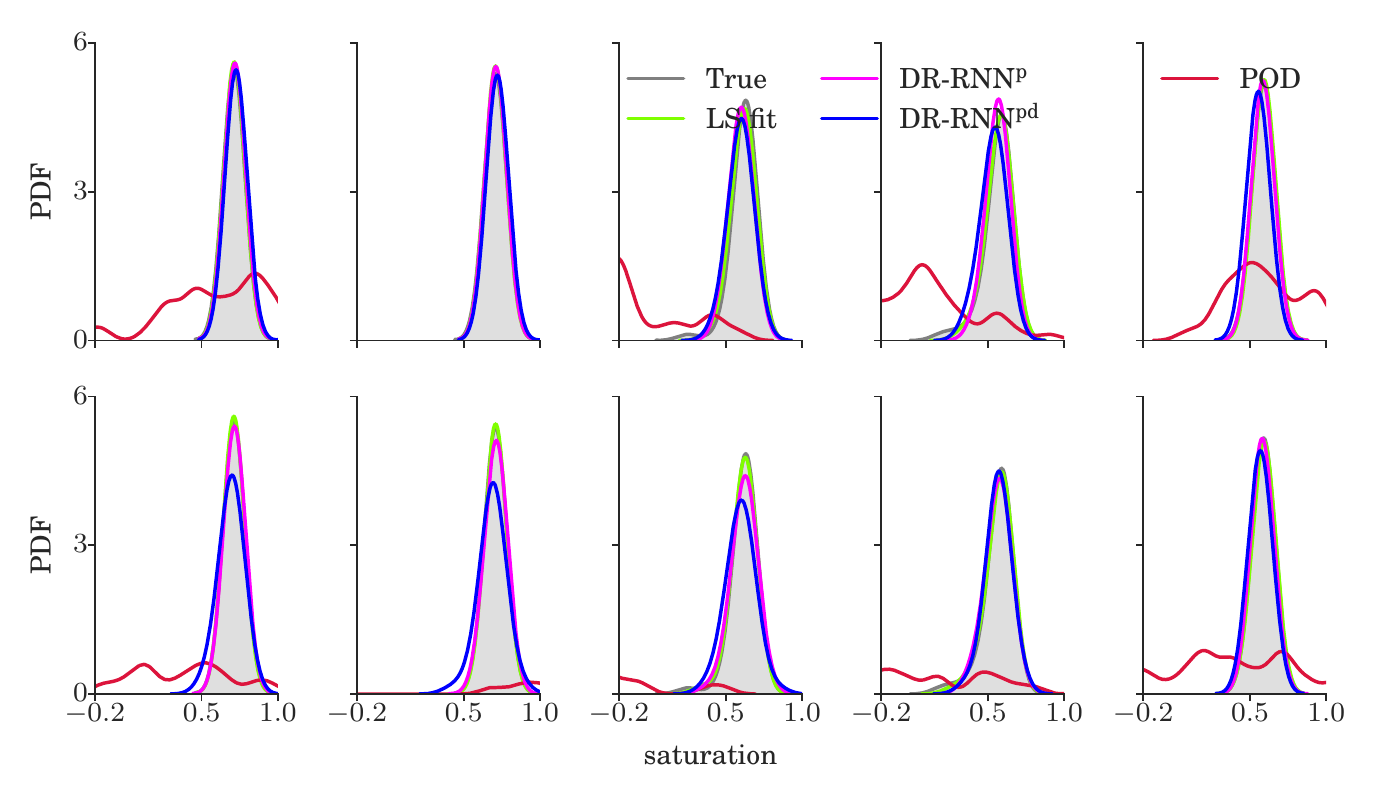}
\caption{Comparison of kernel density estimated probability density function (PDF) at time $= 0.4$ PVI obtained from all ROMs w.r.t. true PDF obtained from the full-order model for test case 2. Top Row: number of POD basis used $=10$. Bottom Row: number of POD basis used $=20$. The plots in each row are arranged as per the numerical notation of the spatial points plotted in Figure~\ref{singularvalues_p2}.}
\label{uniformkdefig}
\end{figure}
%\begin{figure}[H]
%\begin{center}
%\includegraphics{ONEkdecdfplot.pdf}
%%\caption{Left mean square error for the pressure state vector. Right mean square error for the saturation state vector. $r$ denotes number of basis vectors.}
%\end{center}
%\label{msefig}
%\end{figure}
%%%%%%%%%%%%%%%%%%%%%%%%%%%

Figure~\ref{uniformkdefig} compares the saturation PDF estimated from the ensemble of numerical solutions
obtained from all the ROMs and the full model. The plotted results show that DR-RNN$^{\text{p}}$, \mbox{DR-RNN}$^{\text{pd}}$ predictions are nearly indistinguishable from the plots obtained from the full model and are very close to the best possible approximation using LS fit. Further, Figure~\ref{uniformkdefig} shows that all the saturation PDFs obtained from full model are uni-modal distribution. Similar to test case 1, POD reduced model yields inaccurate approximation of the saturation PDFs.

%%%%%%%%%%%%%%%%%%%%%%%%%%%%%%%%%%%%%%%%%%%%%%%%%%%%%%%%%%%%%%%%%%%%%
\begin{table}[h!]
\caption{Performance chart of all the ROMs employed for test case 2. $L_2^{\text{{\tiny rel}}}$ and $L_{2\text{,{\tiny max}}}^{\text{{\tiny rel}}}$ error estimators are defined in the Eq.~\eqref{errorestimators_rel}. The number of POD basis used $=10$ and $20$.}
\centering
% centering table
\begin{tabular}{l c cccc}
\noalign{\smallskip} \hline
%\hline
\noalign{\smallskip
}
Error & \#Basis &\multicolumn{4}{c}{Reduced Order Models} \\
\noalign{\smallskip} \hline %\hline
\noalign{\smallskip
}
% inserts single-line
% Entering 1 st row
 & &LS fit & POD & \mbox{DR-RNN}$^{\text{p}}$ & \mbox{DR-RNN}$^{\text{pd}}$ \\[0.5ex]
 \hline\noalign{\smallskip
}
% Entering 2 nd row
 &10 & 0.09 & 1.30 & 0.10 & 0.12 \\[-1ex]
 \raisebox{1.5ex}{$L_2^{\text{{\tiny rel}}}$} & 20
 & 0.07 & 2.05 & 0.08 & 0.10 \\[1ex]
% Entering 3 rd row
 &10 & 0.19 & 3.5 & 0.21 & 0.22 \\[-1ex]
 \raisebox{1.5ex}{$L_{2\text{,{\tiny max}}}^{\text{{\tiny rel}}} $} & 20
 & 0.16 & 6.2 & 0.18 & 0.22 \\[1ex]
\hline
% inserts single-line
\end{tabular}
\label{uniformtaberror}
\end{table}
%%%%%%%%%%%%%%%%%%%%%%%%%%%%%%%%%%%%%%%%%%%%%%%%%%%%%%%%%%%%%%%%%%%%

We further list in Table~\ref{uniformtaberror}, the error metrics $L_2^{\text{{\tiny rel}}}$ and $L_{2\text{,{\tiny max}}}^{\text{{\tiny rel}}}$ for the saturation fields. From Table~\ref{uniformtaberror}, we can see that the approximation errors obtained from \mbox{DR-RNN}$^{\text{p}}$ and \mbox{DR-RNN}$^{\text{pd}}$ are almost close to the least--squares (best approximation) approximation errors. However, the POD reduced model yields very inaccurate results due to numerical instabilities.

\section{Conclusion}
\label{sec:conclusions}

In this work, we extended the \mbox{DR-RNN} introduced in~\citep{2017nagoor} into nonlinear multi-phase flow problem with distributed uncertain parameters. In this extended formulation, DR-RNN based on the reduced residual obtained from POD-DEIM reduced model is used to construct the reduced order model termed \mbox{DR-RNN}$^{\text{pd}}$. We evaluated the proposed \mbox{DR-RNN}$^{\text{pd}}$ on two forward uncertainty quantification problems involving two-phase flow in subsurface porous media. The uncertainty parameter is the permeability field modeled as log-normal distribution. In the two test cases, full order model and ROMs are solved for $2000$ random permeability realizations to estimate an ensemble based statistics using Monte-Carlo method. full model and POD reduced model used implicit time stepping method as the time step size violates the numerical stability condition. However, \mbox{DR-RNN}$^{\text{pd}}$ architecture employs explicit time stepping procedure for the same step size used in full model and POD reduced model. Hence, \mbox{DR-RNN}$^{\text{pd}}$ had a limited computational complexity $\mathcal{O}(K \times r^2)$ instead of $\mathcal{O}(p \times r^3)$ per saturation update, where $r$ is the dimension of the POD reduced model, $K \ll p$ is the number of stacked network layers in \mbox{DR-RNN} and $p$ is the average number of Newton iterations used in the standard POD-DEIM reduced model. The obtained numerical results shows that \mbox{DR-RNN}$^{\text{pd}}$ provides accurate and stable approximations of the full model in comparison to the standard POD reduced model.

~Future work should consider the development of accurate and stable \mbox{DR-RNN}$^{\text{pd}}$ for UQ tasks involving subsurface flow simulations with the additional effects including the capillary pressure, compressibility, and the gravitational effects. In addition, it will be of interest
to explore the applicability of \mbox{DR-RNN}$^{\text{pd}}$ for UQ tasks with the permeability fields that has randomly oriented channels or barriers. The use of \mbox{DR-RNN}$^{\text{pd}}$ for history matching~\citep{Elsheikh2012,Elsheikh2013-wrr}, where we minimize the mismatch between simulated and field observation data by adjusting the geological model parameters is also expected to show significant reduction of the computational cost.

%%%%\bibliographystyle{plainnat}
%\bibliographystyle{unsrtnat}
%\bibliography{myrefcorrected}

\end{document}